\documentclass[preprint,12pt]{elsarticle}
\biboptions{sort&compress}
\usepackage{array}
\usepackage{longtable,booktabs}
\usepackage{amsmath,bbold,amssymb,epsfig,bm,mathrsfs}
\usepackage{feynmp,color,slashed,nicefrac,exscale,multirow,graphicx}
\usepackage{epstopdf}
\usepackage{longtable}
\usepackage{multirow}

\usepackage[dvipdfm,bookmarks=true,colorlinks,%
            citecolor=blue,linkcolor=blue,hypertex, %
            breaklinks=true]{hyperref}

\journal{Nuclear Physics A }

\begin{document}
\hyphenpenalty=6000
\tolerance=1000
\title{Study of ground state properties of carbon isotopes
       with deformed relativistic Hartree-Bogoliubov theory in continuum}

\author[ITP,UCAS]{Xiang-Xiang Sun}
\author[PCL]{Jie Zhao }
\author[ITP,UCAS,HIRFL,HNNU]{Shan-Gui Zhou \corref{cor1}}
  \ead{sgzhou@itp.ac.cn}
  \cortext[cor1]{Corresponding author.}

\address[ITP]{CAS Key Laboratory of Theoretical Physics,
              Institute of Theoretical Physics,
              Chinese Academy of Sciences, Beijing 100190, China}
\address[UCAS]{School of Physical Sciences,
               University of Chinese Academy of Sciences,
               Beijing 100049, China}
\address[PCL]{Center for Quantum Computing, Peng Cheng Laboratory,
              Shenzhen 518055, China}
\address[HIRFL]{Center of Theoretical Nuclear Physics,
                National Laboratory of Heavy Ion Accelerator,
                Lanzhou 730000, China}
\address[HNNU]{Synergetic Innovation Center for Quantum Effects and Application,
               Hunan Normal University, Changsha 410081, China}

\date{\today}

\begin{abstract}
Ground state properties of carbon isotopes, including root-mean-square radii,
neutron separation energies, single particle spectra, and shapes are systematically
studied with the deformed relativistic Hartree-Bogoliubov theory in continuum.
The calculations with the effective interaction PK1 reproduce
the available data reasonably well.
The shell evolution in this isotopic chain is investigated by examining
the single particle spectra.
The inversion of neutron orbitals $\nu 2s_{1/2}$ and $\nu 1d_{5/2}$ compared with
the order of neutron orbitals in stable nuclei is revealed in
$^{15-20,22}\mathrm{C}$ when these nuclei are constrained to spherical shape.
Neutron halos in $^{15,19,22}$C are studied in detail and their halo structures
are mainly caused by the inversion of ($\nu 2s_{1/2},\nu 1d_{5/2}$) and deformation effects.
In particular, $^{15}$C and $^{22}$C are deformed halo nuclei with
shape decoupling effects in ground states.
\end{abstract}
\maketitle

\section{Introduction}\label{sec:Introduction}

Many interesting and exotic nuclear phenomena have been observed or predicted
in carbon isotopes, including neutron halos, cluster structures, new magicities,
shape decoupling effects, etc.
As far as neutron halos are concerned, $^{15,19}$C are one-neutron halo nuclei
\cite{Bazin1998_PRC57-2156,Fang2004_PRC69-034613} and $^{22}$C
is the heaviest two-neutron halo nucleus with a Borromean structure so far
\cite{Tanaka2010_PRL104-062701,Togano2016_PLB761-412}.
They are all weakly bound, characterized by small one(two)-neutron separation energies,
and have larger radii than neighboring isotopes.
One-neutron separation energies of $^{15}\mathrm{C}$ and $^{19}\mathrm{C}$
are $1.1895 \pm 0.0008$ MeV and $0.5633\pm 0.0915$ MeV in AME2016
\cite{Audi2017_ChinPhysC41-030001,Huang2017_ChinPhysC41-030002,Wang2017_ChinPhysC41-030003} respectively.
The two-neutron separation energy has not been well determined for $^{22}$C up to now.
In 2012, $S_{2n}(^{22}\mathrm{C})$ was deduced to be $-0.14 \pm 0.46$ MeV
from direct mass measurements \cite{Gaudefroy2012_PRL109-202503}.
The empirical value of the two-neutron separation energy for $^{22}$C is
$110\pm 60$ keV in AME2012
\cite{Audi2012_ChinPhysC36-1157,Audi2012_ChinPhysC36-1287,Wang2012_ChinPhysC36-1603}
and
$35\pm20$ keV in AME2016
\cite{Audi2017_ChinPhysC41-030001,Huang2017_ChinPhysC41-030002,Wang2017_ChinPhysC41-030003}.
The matter and proton radii of $^{12-19}\mathrm{C}$ have been reported in
Ref.~\cite{Kanungo2016_PRL117-102501} where the feature of one-neutron halo
in $^{15}\mathrm{C}$ and $^{19}\mathrm{C}$ was illustrated.
The matter radius of $^{22}$C was deduced from two interaction cross section
measurements \cite{Tanaka2010_PRL104-062701,Togano2016_PLB761-412};
the two deduced values differ very much:
$R_m = 5.4 \pm 0.9 $ fm in 2010 \cite{Tanaka2010_PRL104-062701} and
$R_m = 3.44\pm 0.08$ fm in 2016 \cite{Togano2016_PLB761-412}.
More recently, a new value, $R_m(^{22}\mathrm{C}) = 3.38\pm 0.10$ fm,
much smaller than the one given in 2010 \cite{Tanaka2010_PRL104-062701} but
close to the result in 2016,
was obtained by using a Glauber model \cite{Nagahisa2018_PRC97-054614}.

Nuclear halos are formed because of a considerably large occupation
of $s$- or $p$-orbitals close to the threshold of the neutron or proton emission
in weakly bound systems.
Neutron halos in carbon isotopes are mainly caused by the loosely bound
$s$-orbital \cite{Tanihata2013_PPNP68-215,Tanaka2010_PRL104-062701,Togano2016_PLB761-412},
the energy and occupation of which are closely related to
the shell evolution and changes of nuclear magicities \cite{Dobaczewski1994_PRL72-981,Meng1998_PLB419-1,Long2010_PRC81-031302R, Otsuka2020_RMP92-015002}
along this isotopic chain.
A new proton magic number $Z=6$ in $^{13-20}$C was explored
in Ref.~\cite{Tran2018_NatCommun9-1594};
if this is the case,
$^{14}$C would be a doubly-magic nucleus with magic numbers $Z=6$ and $N=8$.
The neutron orbital
$\nu 2s_{1/2}$ is lower than $\nu 1d_{5/2}$ in $^{15}$C,
as pointed out in Ref.~\cite{Tanihata2013_PPNP68-215}.
That is, $\nu 2s_{1/2}$ and $\nu 1d_{5/2}$ are inverted
as compared with single particle levels in stable nuclei.
This inversion results in the formation of a one-neutron halo in this nucleus.
For $^{16,18,20}$C, the ground state configurations are mixtures of
$(\nu 2s_{1/2},\nu 1d_{5/2})$ and
there is no shell gap at $N=14$ in $^{20}$C \cite{Stanoiu2008_PRC78-034315}.
$^{22}$C is the key nucleus to understand whether the shell closure at
$N=16$ \cite{Horiuchi2006_PRC74-034311,Sorlin2008_PPNP61-602,Stanoiu2008_PRC78-034315}
appears or not in this isotopic chain,
which is related to the competition between $\nu 2s_{1/2}$ orbital and $\nu 1d_{5/2}$ orbital \cite{Ozawa2000_PRL84-5493,Otsuka2001_PRL87-082502,
	Cortina-Gil2004_PRL93-062501,Brown2005_PRC72-057301,
	Becheva2006_PRL96-012501,Sorlin2008_PPNP61-602,
	Kanungo2009_PRL102-152501,Hoffman2009_PLB672-17,
	Coraggio2010_PRC81-064303,Kanungo2013_PST152-014002,
	Otsuka2020_RMP92-015002};
this competition also determines the halo configuration in $^{22}$C.
In addition, it is worth mentioning that the
cluster structure has been a very hot topic in carbon isotopes,
such as $3\alpha$ configuration in the Hoyle state of $^{12}$C
\cite{Tohsaki2001_PRL87-192501,Epelbaum2011_PRL106-192501,Epelbaum2012_PRL109-252501,Zhou2016_PRC94-044319},
$3\alpha+xn$ cluster structure in $^{13,14,16}\mathrm{C}$
\cite{Itagaki2004_PRL92-142501,Furutachi2011_PRC83-021303R,
	Suhara2010_PRC82-044301,Baba2014_PRC90-064319,Jansen2014_PRL113-142502,
	Freer2018_RMP90-035004,Tian2016_ChinPhysC40-111001,
	Li2017_PRC95-021303R,Ren2019_SciChinaPMA62-112062},
and rod shapes under high spin and isospin \cite{Zhao2015_PRL115-022501,Ren2019_SciChinaPMA62-112062}.

There have been several systematic theoretical studies on ground
state properties of carbon isotopes.
Calculations made by using a spherical relativistic
Hartree-Bogoliubov (SRHB) theory with the effective interaction NL3 indicated no halos
in carbon isotopes~\cite{Poeschl1997_PRL79-3841}.
With the Fock term included, the spherical
relativistic Hartree-Fock-Bogoliubov (SRHFB) theory
has shown one-neutron halo structure in $^{17,19}$C but no halo
in $^{22}$C \cite{Lu2013_PRC87-034311}.
These SRHB and SRHFB calculations
\cite{Poeschl1997_PRL79-3841,Lu2013_PRC87-034311} were carried out
in coordinate space with spherical symmetry imposed.
Investigations have been made by using covariant density functional theories
\cite{Sharma1999_PRC59-1379,Lalazissis2004_EPJA22-37,Jiang2004_CTP41-79,
	Lu2011_PRC84-014328,Yao2011_PRC84-024306}
or Skyrme density functional theories
\cite{Sagawa2004_PRC70-054316,
	Kanada-Enyo2005_PRC71-014310,Zhang2008_PTP120-129}
with deformation effects considered.
It has been shown that intrinsic deformations in
ground states of carbon isotopes have a strong dependence on the number of neutrons.
There are also some shell model studies
\cite{Coraggio2010_PRC81-064303,Yuan2012_PRC85-064324,Jansen2014_PRL113-142502}
aiming at the description of shell evolution and related properties
in carbon isotopes.

When combining together the halo structure, shell evolution, and deformation effects,
more complex and exotic nuclear phenomenon appears \cite{Zhou2017_PoS-INPC2016-373}.
One typical example is $^{22}$C in which the interplay
among the halo feature, shell evolution, and deformation effects
has been revealed \cite{Sun2018_PLB785-530}
by using the deformed relativistic
Hartree-Bogoliubov theory in continuum (the DRHBc theory).
The DRHBc theory
can self-consistently describe ground state
properties including the halo,
evolution of single particle spectrum, and shell
structure with deformation effects considered
\cite{Zhou2010_PRC82-011301R,Li2012_PRC85-024312,
Li2012_CPL29-042101,Meng2015_JPG42-093101,
Zhang2019_PRC100-034312,Zhang2020_arXiv2001.06599}.
In Ref.~\cite{Sun2018_PLB785-530}, the uncertainties or puzzles
on the matter radius
and the halo configuration of $^{22}$C are resolved and
the ``shrunk" halo feature is attributed
to the inversion of $\nu 2s_{1/2}$ and  $\nu 1d_{5/2}$
and deformation effects.
$^{22}$C was predicted to be a candidate of deformed halo nucleus
with shape decoupling effects, i.e.,
the core and the halo have different shapes.
The shape decoupling in deformed nuclei has been widely investigated by using
the DRHBc theory \cite{Zhou2010_PRC82-011301R,Li2012_PRC85-024312}
and Hartree-Fock-Bogoliubov
theories with the Skyrme interaction
\cite{Pei2013_PRC87-051302R,Pei2014_PRC90-051304R,
	Xiong2016_ChinPhysC40-024101,Wang2017_PRC96-031301R}
and the M3Y-type semirealistic interaction
\cite{Nakada2013_PRC87-014336,Nakada2018_PRC98-011301R}
for axially deformed nuclei.

In this paper we present a systematic study of bulk properties,
including separation energies, matter radius, and deformations,
of this isotopic chain by using the DRHBc theory.
We focus on halo structures, deformation effects, and the evolution of
single neutron levels with quadrupole deformation.
The paper is organized as follows.
The main formulae of the DRHBc theory are displayed in Sec.~\ref{sec:formulae}.
The results and discussions of bulk properties of carbon isotopes are given in Sec.~\ref{sec:bulk}.
The structure of odd $A$ nuclei $^{15,17,19}$C is discussed in Sec.~\ref{sec:C15-17-19}.
The results for $^{22}$C from DRHBc calculations are shown in Sec.~\ref{sec:C22}.
In Sec.~\ref{sec:summary} we summarize this work.

\section{The DRHBc theory}\label{sec:formulae}

The covariant density function theory (CDFT) is one of the most successful models
for the study of nuclear structure in almost the whole nuclear landscape
\cite{Serot1986_ANP16-1,Reinhard1989_RPP52-439,
Ring1996_PPNP37-193,Vretenar2005_PR409-101,Meng2006_PPNP57-470,Niksic2011_PPNP66-519,
Meng2013_FrontPhys8-55,Liang2015_PR570-1,Meng2015_JPG42-093101,Meng2016_RDFNS}.
The relativistic continuum Hartree-Bogoliubov (RCHB) theory
\cite{Meng1996_PRL77-3963,Meng1998_NPA635-3} and
the DRHBc theory
\cite{Zhou2010_PRC82-011301R,Li2012_PRC85-024312} have been developed and used to
study ground state properties of spherical and deformed halo nuclei, respectively.
The detailed formulae for the DRHBc theory can be found in
Refs.~\cite{Zhou2010_PRC82-011301R,Li2012_PRC85-024312,Li2012_CPL29-042101,Chen2012_PRC85-067301}.
Here we give briefly the main ones for convenience of
discussions in the following sections.
The relativistic Hartree-Bogoliubov (RHB) equation for nucleons \cite{Kucharek1991_ZPA339-23} reads,
\begin{equation}\label{equ.1}
\left(
  \begin{array}{cc}
   h_D - \lambda_\tau &    \Delta     \\
   -\Delta^*    & -h^*_D+\lambda_\tau \\
  \end{array}
\right)
 \left( { U_{k} \atop V_{k} } \right)
 =
 E_{k}
 \left( { U_k \atop V_{k} }  \right),
\end{equation}
which is solved in a spherical Dirac Woods-Saxon (WS) basis \cite{Zhou2003_PRC68-034323}.
$E_k$ is the quasi particle energy, $\lambda_\tau$ $(\tau=\mathrm{n,p})$ is the chemical potential,
and $U_k$ and $V_k$ are quasi particle wave functions.
$h_\mathrm{D}$ is the Dirac Hamiltonian
\begin{equation}
  h_\mathrm{D} = \bm \alpha\cdot \bm p + V(\bm r) + \beta[M+S(\bm r)],
\end{equation}
where $S(\bm r)$ and $V(\bm r)$ are scalar and vector potentials.
The pairing potential reads
\begin{equation}
  \Delta(\bm r_1, \bm r_2) = V^{pp}(\bm r_1, \bm r_2) \kappa(\bm r_1, \bm r_2),
\end{equation}
where we use a density dependent zero-range force in the particle-particle ($pp$) channel,
\begin{equation}
  V^{pp}(\bm r_1, \bm r_2)= \frac{1}{2}V_0(1-p^\sigma)
  \delta(\bm r_1, \bm r_2)
  \left(1-\frac{\rho({\bm r_1})}{\rho_{\mathrm{sat}}}\right),
 \label{eq:pairing-force}
\end{equation}
and $\kappa(\bm{r}_1,\bm{r}_2)$ is the
pairing tensor \cite{Ring1980,Blaizot1985_QTFS}.

For axially deformed nuclei with spatial reflection symmetry,
we expand scalar densities and potentials in terms of Legendre
polynomials,
\begin{equation}
  f(\bm r) = \sum_\lambda f_\lambda (r)P_\lambda (\cos\theta),\quad \lambda = 0,2,4,\cdots,
  \label{eq:expansion}
\end{equation}
with
\begin{equation}
  f_\lambda (r) = \frac{2\lambda +1}{4\pi}\int d\Omega f(\bm r )P_\lambda (\cos\theta).
\end{equation}

The ground state of an even-even nucleus is obtained by solving the RHB
equation iteratively. For systems with odd number of neutrons (protons),
the ground state is obtained by taking
the blocking effect into account. The details about how to deal with the blocking effect
in the DRHBc theory can be found in
Ref.~\cite{Li2012_CPL29-042101}.

In order to get the potential energy curve and examine the evolution of single particle levels
with deformation, one can make constraint calculations \cite{Ring1980}.
The augmented Lagrangian method \cite{Staszczak2010_EPJA46-85}
has been implemented in the DRHBc theory
and the constrained Dirac Hamiltonian $h'_\mathrm{D}$ is written as
\begin{equation}
  h'_\mathrm{D} = h_\mathrm{D} + c_1 \left(\langle \hat Q_{2} \rangle -\bar Q_{2}  \right)
   + c_2 \left(\langle \hat Q_{2} \rangle -\bar Q_{2} \right)^2,
\end{equation}
where $c_1$ is the Lagrange multiplier,
$c_2$ is the penalty parameter, and $\bar Q_{2}$ is the desired expectation value
of the quadrupole moment $\hat{Q}_2$.
In the iteration, $c_2$ is kept as a constant
and $c_1$ in the $m$th step is re-adjusted in the following way,
\begin{equation}
  c_1^{(m)} = c_1^{(m-1)} - 2c_2 \left(\langle \hat Q_{2} \rangle^{(m-1)} -\bar Q_{2}  \right).
\end{equation}

The bulk properties of a nucleus in question,
including the binding energy, radius, deformation,
etc., can be calculated from quasi-particle wave functions and densities \cite{Li2012_PRC85-024312}.
The densities are important for the study of halo nuclei.
Next we briefly discuss how to calculate
the densities with emphasis on odd $A$ systems.

The halo is mainly determined
by occupation probabilities and wave functions of weakly bound orbitals
\cite{Chen2014_PRC89-014312}.
For odd $A$ nuclei,
the occupation probability of the blocked level is around 0.5.
%
The neutron (proton) density of an odd $N(Z)$ nucleus is
calculated as \cite{Li2012_CPL29-042101}
\begin{equation}
\rho(\bm r) = \sum_{k \neq k_b} |V_k(\bm r)|^2 + |U_{k_b}(\bm r)|^2.
\end{equation}
The density of a halo nucleus extends far away from the center of the nucleus, which is determined by the asymptotic behavior of the wave function of the valence nucleon(s).
For a bound system, the Fermi surface $\lambda_\tau < 0$ and the
asymptotic behavior of quasi particle wave functions $V_k(r)$ and $U_k(r)$
are derived as \cite{Dobaczewski1984_NPA422-103,Chen2014_PRC89-014312}
\begin{align}\label{Eq:wf}
  & V_k(r) \propto \frac{\exp(-\kappa_{k+} r)}{r}, \\
  & U_k(r) \propto \frac{\exp(-\kappa_{k-} r)}{r},
\end{align}
where $\kappa_{k\pm} = \sqrt{2m(|\lambda_\tau| \pm E_k)}$,
$E_k = \sqrt{(\varepsilon_k - \lambda_\tau)^2 + \Delta_k^2}$ is the quasi particle
energy, and $\varepsilon_k$ and $\Delta_k$ denote the single particle energy and
pairing gap. $\Delta_k $ stays finite or equals zero,
so $|U_k(r)|^2$ diminishes more slowly to zero with $r$ than $|V_k(r)|^2$.
Therefore,
the asymptotic behavior of $U_k(\bm r)$
leads to a more extended density distribution \cite{Nakada2018_PRC98-011301R}
for odd $A$ nuclei,
thus enhancing the formation of a halo.

\section{Bulk properties of carbon isotopes}\label{sec:bulk}

We have investigated bulk properties of carbon isotopes by using the DRHBc theory.
The two commonly used and successful effective interactions
PK1 \cite{Long2004_PRC69-034319} and NL3
\cite{Lalazissis1997_PRC55-540} are adopted in the particle-hole
($ph$) channel while a density dependent zero-range pairing force
(\ref{eq:pairing-force}) 
is used in the $pp$ channel.
In this work, the Dirac WS basis is generated in a box
with the size $R_\mathrm{box}$ fixed to be 25 fm and the
mesh size $\Delta r =0.1$ fm.
An energy cutoff $E_{\rm{cut}}^+ =100$ MeV is made to
determine the Dirac WS basis for positive-energy states
in the Fermi sea
and the number of negative-energy
states in the Dirac sea is taken the same as that of positive-energy
states \cite{Zhou2003_PRC68-034323,Li2012_PRC85-024312}.
A cutoff on the angular momentum is made up to
$J_\mathrm{max} = \frac{21}{2}\hbar$.
The pairing strength $V_0$ and cut-off energy
$E_{\mathrm{cut}}^{\mathrm{q.p}}$ in
the quasi-particle space are 355 MeV$\cdot$fm$^{3}$
and 60 MeV respectively, which are the same as in Ref.~\cite{Sun2018_PLB785-530}.
The maximal order $\lambda_\mathrm{max}$ in the Legendre expansion (\ref{eq:expansion})
is taken to be 4.
The ground states of odd $A$ carbon isotopes
are calculated with blocking effects considered.
For each odd-$A$ nucleus, we block several single particle levels around the neutron Fermi surface
and take the one with maximal binding energy as the ground state.

\begin{center}
\begin{longtable}{cllr@{.}llr@{.}l}
\caption{Experimental and calculated neutron
separation energy ($S_{2n}$ for even $A$ and $S_n$ for odd $A$ nuclei),
rms matter radius $R_m$, and
quadrupole deformation parameter $\beta_2$.
$\Omega^\pi$ represents the blocked
level for each odd $A$ nucleus.
Experimental (Expt.) values of $S_{2n}$ $(S_n)$, $R_m$,
and $\beta_2$
are taken from Refs.~\cite{Wang2017_ChinPhysC41-030003,
Kanungo2016_PRL117-102501,Pritychenko2016_ADNDT107-1}, respectively,
except otherwise noted.
}
\label{table:C_bulk}

\\
\toprule[1pt]
	  & $\Omega^\pi$ &     & \multicolumn{2}{c}{$S_{2n}$ ($S_n$) (MeV)} & $R_m$ (fm) & \multicolumn{2}{c}{\quad$\beta_2$} \\ \hline
\endfirsthead
\hline
	  & $\Omega^\pi$ &    & \multicolumn{2}{c}{$S_{2n}$ ($S_n$) (MeV)} & $R_m$ (fm) & \multicolumn{2}{c}{\quad$\beta_2$} \\ \hline
\endhead

\hline
\endfoot
	$^{12}$C& &Expt.& 31&0950(1)&2.35(2)&0&583(15) \\
	        & &     & \multicolumn{2}{c}{}  &       &$-$0&40(2)~\cite{Yasue1983_NPA394-29}\\
             & &PK1  & 31&1020   & 2.37  &$-$0&3223\\
             & &NL3  & 30&1811   & 2.25  &0&0000\\
    \hline
    $^{13}$C& &Expt.& 4&8304&2.28(4)\\
            &$1/2^-$&PK1  & 7&7551   &2.41     &0&0000 \\
            &$1/2^-$&NL3  & 7&5501   &2.40     &0&0000\\
        \hline
    $^{14}$C&&Expt.&  12&8152&2.33(7) \\
            &&PK1  &  16&8611   &2.48    &0&0000\\
            &&NL3  &  16&5401   &2.49    &0&0000\\\hline
    $^{15}$C&&Expt.&  1&1895(8)&2.54(4)  \\
            &$1/2^+$&PK1  &1&2162 &2.64&0&2559  \\
            &$1/2^+$&NL3  &1&5832  &2.65&0&2510  \\
            &$5/2^+$&PK1  &0&9584 &2.60&$-0$&$1746$  \\
            &$5/2^+$&NL3  &1&1309 &2.62&$-0$&$1647$  \\
            \hline
    $^{16}$C&&Expt.& 5&3402(35)&2.74(3) &0&323(18) \\
            &&     &  \multicolumn{2}{c}{}      &        &0&$356^{+0.25}_{-0.23}$~\cite{Jiang2020_PRC101-024601} \\
            &&PK1  & 4&4421    &2.70    &$-0$&$2260$ \\
            &&NL3  & 5&8507    &2.71    &$-0$&$1727$ \\\hline
    $^{17}$C&&Expt.& 0&7164(137)&2.76(3)&0&52(4)~\cite{Elekes2005_PLB614-174} \\
            &$3/2^+$&PK1  &1&9631&2.81&0&4621  \\
            &$3/2^+$&NL3  &1&3016&2.83&0&4611\\
            &$1/2^+$&PK1  &1&4245&2.85&$-0$&$3279$  \\
            &$1/2^+$&NL3  &1&3316&2.86&$-0$&$2886$ \\
            \hline
    $^{18}$C&&Expt.& 4&8022(295)&2.86(4)&0&289($^{+20}_{-13}$)  \\
            &&PK1  & 5&2768     &2.89   &$-0$&$3763$ \\
            &&NL3  & 5&7614     &2.90   &$-0$&$3514$ \\
            \hline
    $^{19}$C&&Expt.& 0&5633(915)&3.16(7)& 0&29(3)~\cite{Elekes2005_PLB614-174} \\
            & &    & \multicolumn{2}{c}{}&3.10($^{+5}_{-3}$)~\cite{Togano2016_PLB761-412} \\
            &$1/2^+$&PK1  &0&0304&3.05&0&3743   \\
            &$3/2^+$&PK1  &1&6046&3.02&$-$0&4459 \\
            &$3/2^+$&NL3  &0&9845&3.02&$-$0&4395  \\
            \hline
    $^{20}$C&&Expt.& 2&9116(2449)&2.98(5)~\cite{Togano2016_PLB761-412}&0&405($^{+89}_{-45}$) \\
            &&PK1  & 2&9014     &3.07 &$-$0&4661   \\
            &&NL3  & 4&3989     &3.07 &$-$0&4098 \\
            \hline
    $^{22}$C&&Expt.& 0&1006(6259)&3.44(8)~\cite{Togano2016_PLB761-412} \\
            &&     &$-0$&14(46)~\cite{Gaudefroy2012_PRL109-202503} &5.4(9)~\cite{Tanaka2010_PRL104-062701} \\
            &&     &\multicolumn{2}{c}{}&3.38(10)~\cite{Nagahisa2018_PRC97-054614} \\
            &&PK1  & 0&3998      &3.25&$-$0&2649 \\
            &&NL3  & 1&9343      &3.22&$-$0&2311 \\
\bottomrule[1pt]
\end{longtable}
\label{symbols}
\end{center}

\begin{figure}[htbp]
	\begin{center}
		\includegraphics[width=0.6\textwidth]{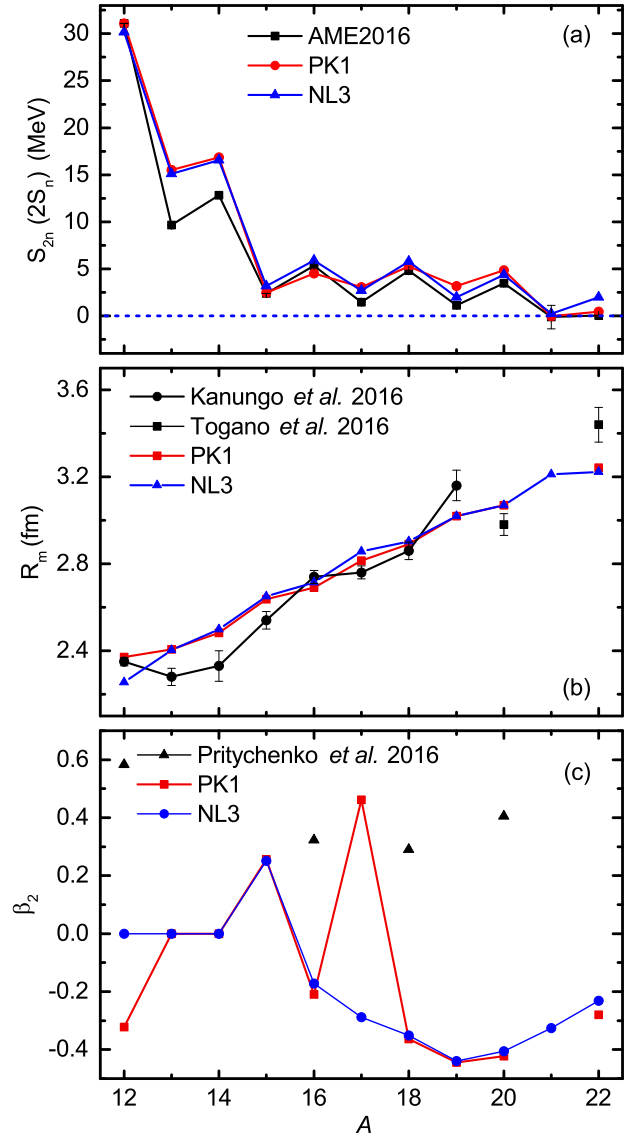}
	\end{center}
	\caption{(Color online) (a) Neutrons separation energy $S_{2n}$ (for even $A$ nuclei) or $2S_n$ (for odd $A$ nuclei), (b) rms matter radius $R_m$,
		and (c) quadrupole deformation parameter $\beta_2$ for carbon isotopes.}
	\label{figure:Bulk}
\end{figure}

Experimental and calculated neutron separation energies, rms matter radii $R_m$,
and quadrupole deformation parameters $\beta_2$ of carbon isotopes are presented
in Table~\ref{table:C_bulk}.
For each of these nuclei, there are usually more than one minimum in
the potential energy curve.
For spherical (deformed) odd $A$ nuclei, we show the results with
blocking one (two) single neutron level(s)
around the Fermi surface.
In Table~\ref{table:C_bulk}, one can see the spin and parity of
the ground state of $^{15}$C ($^{19}$C) is $1/2^+$ (${3/2}^+$).
For $^{17}$C, the ground state spin-parity is $J^\pi = 3/2^+$ with PK1 and $J^\pi = 1/2^+$ with NL3.
Next we give detailed discussions about these results.

In Fig.~\ref{figure:Bulk}(a), the calculated $S_{2n}$
for even $A$ nuclei and $2S_n$ for odd $A$
nuclei are shown and compared with the
data taken from Ref.~\cite{Wang2017_ChinPhysC41-030003}.
The odd-even mass staggering
in carbon isotopes can be reproduced by calculations with both parameter sets.
The calculated $S_{n}$ for $^{13}\rm C$ and $S_{2n}$
for $^{14}\rm C$
are larger than experimental values.
For other carbon isotopes, calculations reproduce
the experiment reasonably well.
$^{22}\mathrm{C}$ is the heaviest Borromean nucleus observed so far.
Our calculations can reproduce the Borromean feature \cite{Zhukov1993_PR231-151}
of $^{22}$C with PK1 but not NL3:
With both effective interactions, $^{20,22}$C are bound;
$^{21}$C is unbound with PK1
($S_n = -28$ keV) but bound with NL3 ($S_n = 0.22$ MeV).
As mentioned in the Introduction, the separation energy $S_{2n}$ has
not been well determined for $^{22}$C yet:
$110\pm60$ keV \cite{Wang2012_ChinPhysC36-1603},
$-140\pm460$ keV \cite{Gaudefroy2012_PRL109-202503},
and $35\pm20$ keV \cite{Wang2017_ChinPhysC41-030003}.
The calculated $S_{2n}$ of $^{22}$C is 0.40 MeV with PK1 and 1.93 MeV with NL3.

Experimental and calculated rms matter radii $R_m$ of carbon isotopes
are shown in Fig.~\ref{figure:Bulk}(b).
The experimental $R_m$ values of $^{12-19}\mathrm{C}$
and $^{20,22}$C are taken from
Refs.~\cite{Kanungo2016_PRL117-102501,Togano2016_PLB761-412} respectively.
The recent progress on the radii of carbon isotopes has been summarized
in Ref.~\cite{Fortune2016_PRC94-064307}.
The large enhancement of $R_m$ with $A$ is usually regarded as a sign of a halo.
One can find such signs at $A=15,19$, and $22$ in Fig.~\ref{figure:Bulk}(b) obviously.
There are slightly sudden increases in the calculated $R_m$ at $A=15$ and $A=19$.
The calculated $R_m$ for $^{15}\mathrm{C}$ with PK1
and NL3 are 2.64 fm and 2.65 fm, a bit larger than the experimental value $2.54\pm0.04$ fm \cite{Kanungo2016_PRL117-102501}.
$R_m$ of $^{19}$C calculated with PK1 and NL3 are both $3.02$ fm,
which is smaller than experimental values $3.10^{+0.05}_{-0.03}$ fm \cite{Togano2016_PLB761-412}
and $3.16 \pm 0.07$ fm \cite{Kanungo2016_PRL117-102501}.
For $^{20}\mathrm{C}$, the $R_m$ values calculated with PK1 and NL3 are both 3.07 fm,
slightly larger than
the experimental value $2.97^{+0.03}_{-0.05}$ fm \cite{Togano2016_PLB761-412}.
%
%
According to an early experiment,
$^{22}$C should be a huge halo nucleus
with $R_m= 5.4 \pm 0.9$ fm \cite{Tanaka2010_PRL104-062701}.
However, much smaller $R_m$ values were
deduced from a more recent measurement of
interaction cross sections, $R_m=3.44\pm 0.08$ fm \cite{Togano2016_PLB761-412}
and $3.38\pm 0.10$ fm \cite{Nagahisa2018_PRC97-054614}.
This means the halo in $^{22}$C was ``shrunk".
Our calculated $R_m$ for $^{22}$C, 3.25 (3.22) fm with PK1 (NL3),
is close to these recent experimental results.
The mechanism of the shrinkage has been proposed
in Ref.~\cite{Sun2018_PLB785-530}
and will also be discussed in Sec.~\ref{sec:C22}.


The quadrupole deformation parameters $\beta_2$ in ground states of nuclei
in carbon isotopes are shown in Fig.~\ref{figure:Bulk}(c).
Generally speaking, calculated values of $\beta_2$ agree with available experimental ones well.
In our calculations, the ground state of $^{12}$C is oblate with PK1
but spherical with NL3.
The oblate shape of $^{12}\mathrm{C}$ with PK1
is consistent with an inelastic
scattering experiment \cite{Yasue1983_NPA394-29}.
A recent experiment \cite{Kumar-Raju2018_PLB777-250} confirms
the oblate shape of $^{12}$C in its ground state as well.
Spherical $^{13,14}\mathrm{C}$, prolate
$^{15}$C, and oblate $^{16,18-20,22}\mathrm{C}$
are obtained in calculations with both PK1 and NL3.
In a recent work, Jiang et al. studied the quadrupole
deformation of $^{16}\mathrm{C}$ by performing proton
and deuteron inelastic scattering experiments
\cite{Jiang2020_PRC101-024601}.
From the $B(E2; 2^+_1 \rightarrow 0^+_\mathrm{g.s.})$
value (4.34$^{+2.27}_{-1.85}$ e$^2$ fm$^4$)
given in Ref.~\cite{Jiang2020_PRC101-024601}
one can deduce $\beta_2 = 0.356^{+0.25}_{-0.23}$ for $^{16}\mathrm{C}$.
Different shapes are predicted for $^{17}$C in calculations with PK1 (prolate) and NL3 (oblate).
Calculations by using Skyrme density functional SkO' for $^{17}$C
predicted a prolate shape in the ground state \cite{Sagawa2008_PRC78-041304R}
with the spin-parity $J^\pi = 3/2^+$, which agrees with our calculations with PK1.
The $\beta_2$ values of $^{17}$C and $^{19}$C
are deduced to be $0.52\pm0.04$
and $0.29\pm0.03$ in Ref.~\cite{Elekes2005_PLB614-174}.
The intrinsic shape of $^{16,18}\mathrm{C}$
is prolate and $^{12,22}\mathrm{C}$ show an oblate shape
in the calculation with the MDC-RMF model \cite{Lu2011_PRC84-014328},
which agrees with Ref.~\cite{Sagawa2004_PRC70-054316}.
The $\beta_2$ of $^{22}\mathrm{C}$
with PK1 and NL3
is close to $-0.25$,
which is similar to results in Refs.~\cite{Sagawa2004_PRC70-054316,Lu2011_PRC84-014328}.

\begin{figure*}[htb]
\begin{center}
\includegraphics[width=\textwidth]{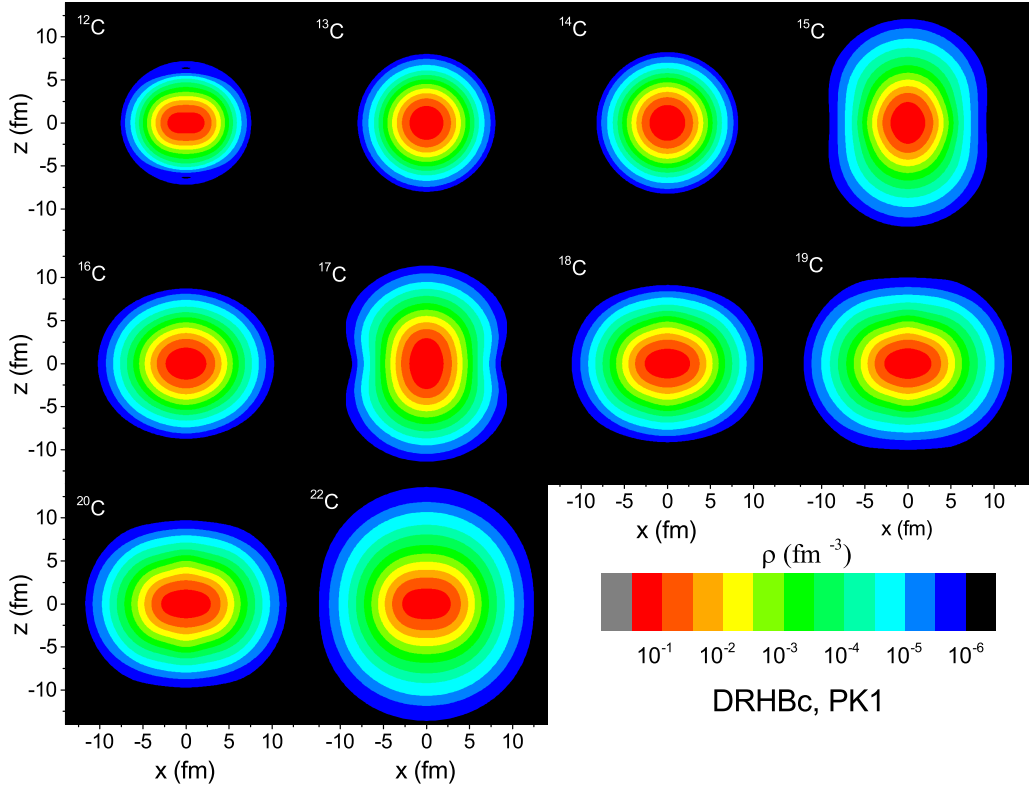}
\end{center}
\caption{(Color online)
Two-dimensional neutron density distributions for $^{12-20,22}$C
from DRHBc calculations with PK1.
The $z$-axis is the symmetry axis for axially deformed systems.
}
\label{fig:density-C}
\end{figure*}

From the above discussions on neutron separation energies,
rms matter radii, and deformations for carbon isotopes,
it can be seen that the effective interaction PK1 can give a proper description for carbon isotopes.
Especially, the Borromean feature of $^{22}$C can be reproduced with PK1.
Next we will focus
on the results obtained with the parameter set PK1.

In Fig.~\ref{fig:density-C},
the two-dimensional neutron density
distributions for the ground states of nuclei in this isotopic chain calculated with PK1 are presented.
The shape evolution with $A$ can be clearly seen from these density profiles.
Density distributions of $^{15}$C and $^{22}$C have larger spatial extensions than adjacent isotopes.

\begin{figure*}[htbp]
\begin{center}
\includegraphics[width=\textwidth]{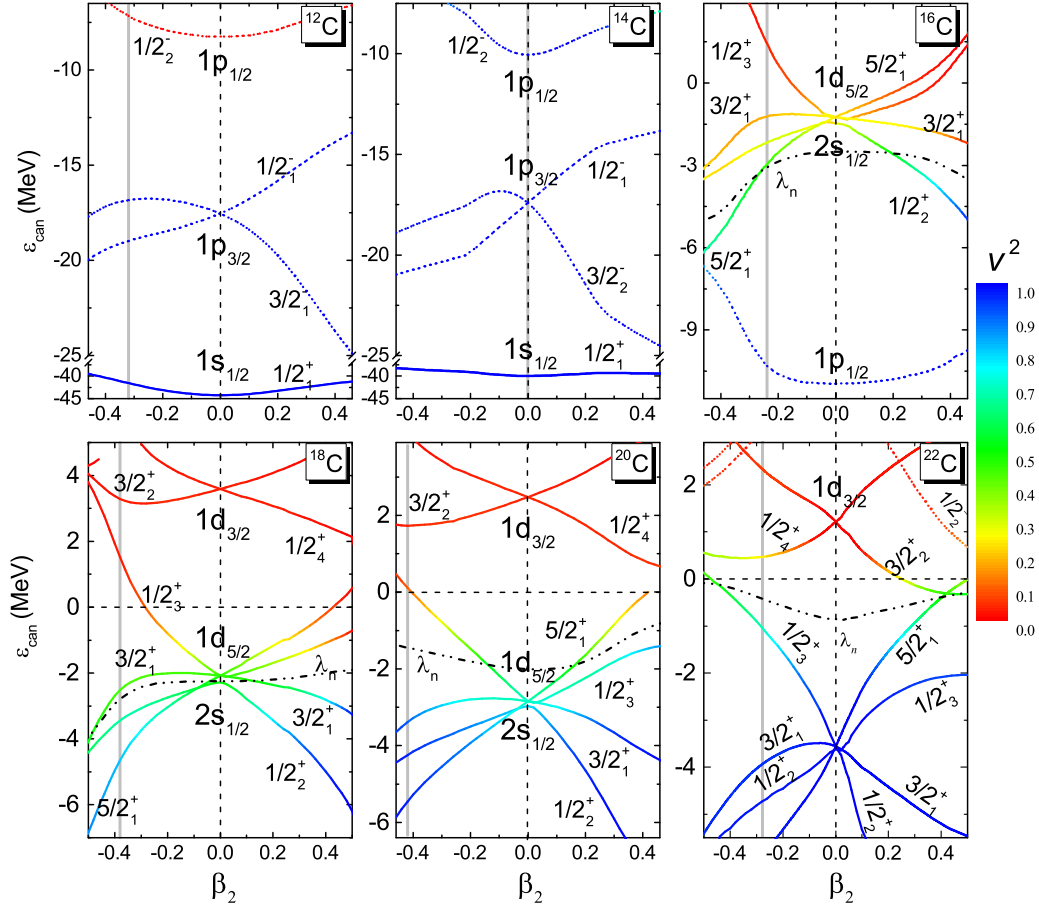}
\end{center}
\caption{(Color online)
Single neutron levels in the canonical basis of $^{12,14,16,18,20,22}$C
from deformation constraint DRHBc calculations with PK1.
Levels with positive (negative) parity are presented by solid (dotted) lines.
Each level is labeled with $\Omega^\pi_i$ where $\pi$ is the parity,
$\Omega$ is the projection of the angular momentum on the symmetry axis, and
$i$ is used to order the level in each $\Omega^\pi$-block.
The occupation probability $\nu^2$ of each orbital
is represented with different colors.
The ground state is indicated by the gray vertical line.
The Fermi levels ($\lambda_n$) are displayed by black dash-dotted lines.
For $^{12}$C and $^{14}$C, all single particle levels shown in the figures
are fully occupied and the Fermi levels are not shown.
}
\label{figure:SPL-iso}
\end{figure*}

Next we discuss the shell evolution in this isotopic chain by investigating single particle spectra.
Single neutron spectra for carbon isotopes with even $A$ are shown in Fig. \ref{figure:SPL-iso}.
For each nucleus, quadrupole deformation constraint calculations are carried out to get the
single neutron levels versus the deformation parameter $\beta_2$.
In the spherical case, the neutron orbital $\nu 2s_{1/2}$ is lower than
$\nu 1d_{5/2}$ for $^{16,18,20,22}$C,
which means $\nu 2s_{1/2}$ and $\nu 1d_{5/2}$ are inverted
compared with the order of neutron orbitals in stable nuclei.
Furthermore,
$\nu 2s_{1/2}$ and $\nu 1d_{5/2}$ are nearly degenerate
in these carbon isotopes.
The inversion and near degeneracy of ($\nu 2s_{1/2}$, $\nu 1d_{5/2}$) lead to the appearance of the magic number $N=16$, corresponding to the large gap between $\nu 1d_{3/2}$ and $\nu 1d_{5/2}$.
This inversion in carbon isotopes has also been predicted by using a shell model
approach with chiral $NN$ potential \cite{Coraggio2010_PRC81-064303}.
The shell closure at $N=14$
\cite{Stanoiu2008_PRC78-034315,Strongman2009_PRC80-021302R,Jansen2014_PRL113-142502}
no longer exists in DRHBc calculations with PK1 due to the near
degeneracy of ($\nu 2s_{1/2}, \nu 1d_{5/2}$).
When the deformation comes in, it is more complicated to identify
new magic numbers according to single particle levels (SPLs)
because of the large level density
caused by quadrupole correlations.
In $^{16,18,20,22}$C,
$s$- and $d$-wave orbitals mix strongly due to quadrupole correlations, resulting in deformed ground states.
In particular, the magic number $N=16$, which appears in the spherical limit, disappears
after taking deformation effects into account in $^{22}$C \cite{Sun2018_PLB785-530}.
The SPLs and configurations of odd $A$ nuclei
$^{15,17,19}$C will be shown and
discussed one by one in the next section.

\section{Structure of  $^{15,17}$C and $^{19}$C}
\label{sec:C15-17-19}

The one-neutron halo structure in $^{15}\mathrm{C}$ and $^{19}$C has been confirmed
experimentally
\cite{Bazin1998_PRC57-2156,Pramanik2003_PLB551-63}.
$^{17}\mathrm{C}$ is not a halo nucleus in the ground state.
The experimental results,
including the spin-parity $J^\pi$ of the ground state
and valence orbitals for these nuclei, are listed in Table~\ref{tab:C15-17-19}.
More experimental information can be found in Ref.~\cite{Tanihata2013_PPNP68-215}.


\begin{table}[htb]
\centering
\caption{
	Experimental ground state spin-parity $J^\pi$ and valence orbitals of
	$^{15,17,19}$C.
	}	\label{tab:C15-17-19}
        \begin{tabular}{p{0.22\textwidth}p{0.35\textwidth}p{0.3\textwidth}<{\centering}}
\toprule[1pt]
			Nucleus                & ${J}^\pi$ (g.s.) &  Valence orbital  \\
			\hline
			$^{15}\mathrm{C}$      & $1/2^+$  \cite{Murillo1994_NPA579-125,Terry2004_PRC69-054306,Fang2004_PRC69-034613}              & $s$-wave       \\
			$^{17}\mathrm{C}$      & $3/2^+$  \cite{Maddalena2001_PRC63-024613,Elekes2005_PLB614-174,Ueno2013_PRC87-034316}
			& $d$-wave      \\
			$^{19}\mathrm{C}$      & $1/2^+$  \cite{Nakamura1999_PRL83-1112,Maddalena2001_PRC63-024613,Elekes2005_PLB614-174,Hwang2017_PLB769-503}
			& $s$-wave    \\
        \bottomrule[1pt]
	\end{tabular}
\end{table}

\begin{figure}[htbp]
\begin{center}
\includegraphics[width=0.8\textwidth]{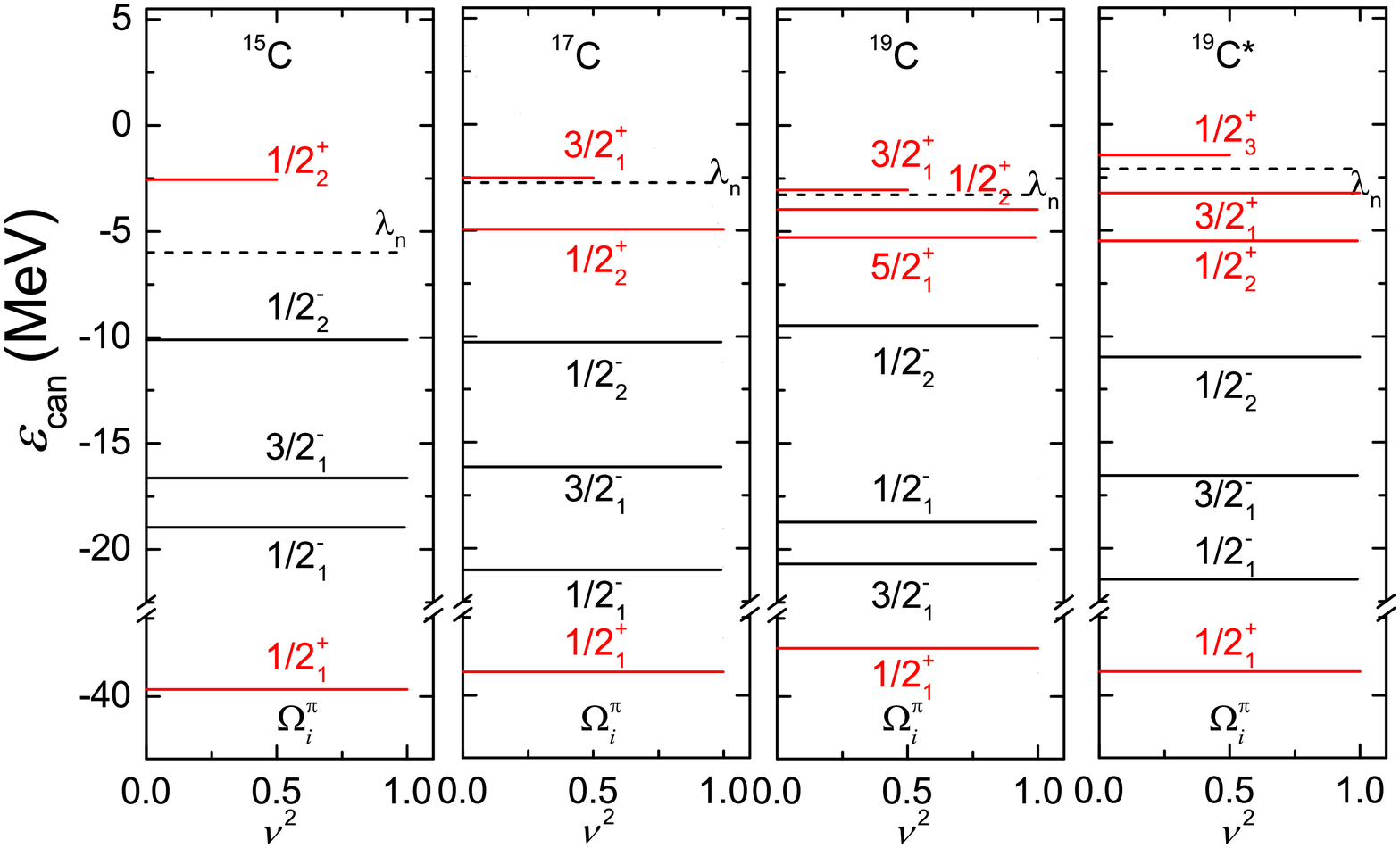}
\end{center}
\caption{(Color online)
Single neutron levels in the canonical basis of $^{15,17,19}$C and $^{19}\mathrm{C}^*$
from DRHBc calculations with PK1.
Each level is labeled with $\Omega^\pi_i$
where $\pi$ is the parity,
$\Omega$ is the projection of the angular momentum on the symmetry axis, and
$i$ is used to order the level in each $\Omega^\pi$-block.
The length of each level is proportional to the occupation probability $v^2$.
Levels with positive (negative) parity are presented by red (black) lines.
The dash line represents the neutron Fermi surface.
{\color{red}}
}
\label{fig:SPL-C15-17-19}
\end{figure}

Halo nuclei are weakly bound and well associated with pairing correlations
and the contribution from the continuum above the threshold of
particle emission
\cite{Hansen1987_EPL4-409,Dobaczewski1996_PRC53-2809,
    Meng1996_PRL77-3963,Meng1998_PRL80-460,Meng1998_NPA635-3,
	Jensen2004_RMP76-215,
	Tanihata2013_PPNP68-215,
	Riisager2013_PST152-014001,
	Zhang2014_PLB730-30,Zhang2017_PRC95-014316}.
When valence neutron(s) occupy low $l$-orbitals ($s$- or $p$-wave)
close to the threshold,
including weakly bound ones and those embedded in the continuum,
a halo appears.
Single neutron levels of $^{15,17,19}$C and $^{19}$C$^*$
(the halo state corresponding to the second energy minimum in the potential
energy surface with the single neutron level $1/2^+_3$ blocked)
in the canonical basis are given in Fig.~\ref{fig:SPL-C15-17-19}.
In our DRHBc calculations with blocking effects taken into account,
pairing gaps of $^{15,17,19}$C and $^{19}\mathrm{C}^*$
almost vanish and continuum states are empty.
Therefore, whether a halo emerges or not
in these nuclei is totally determined by the
configuration of the weakly bound level and blocked orbitals.

\subsection{$^{15}\mathrm{C}$}\label{subsec:C-15}

\begin{figure}[htb]
\begin{center}
\includegraphics[width=0.7\textwidth]{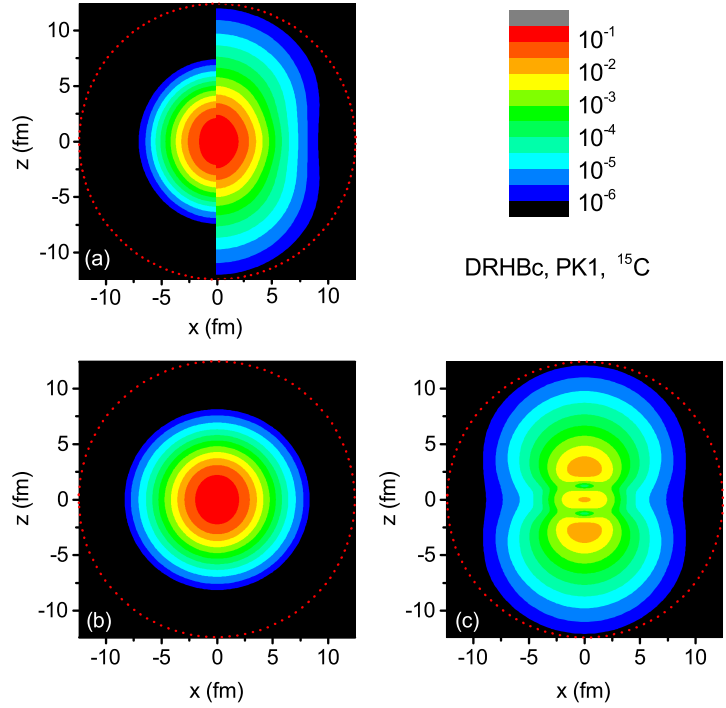}
\end{center}
\caption{(Color online)
Density profiles of $^{15}$C from DRHBc calculations with PK1.
The $z$-axis is the symmetry axis.
(a) The proton ($x<0$) and neutron ($x>0$) density profiles,
(b) the density profile of the neutron core, and
(c) the density profile of the neutron halo.
In each plot, a dotted circle is drawn to guide the eye.
}
\label{fig:den-C15-PK1}
\end{figure}

In Fig.~\ref{fig:den-C15-PK1}, density profiles of $^{15}$C obtained from DRHBc
calculations with PK1 are presented. The density distributions of protons
and neutrons are shown in the left and right parts of
Fig.~\ref{fig:den-C15-PK1}(a), respectively.
It is obvious that the neutron density extends much farther than protons,
hinting a neutron halo in $^{15}$C.

As Fig.~\ref{fig:SPL-C15-17-19} shows,
the valence neutron of $^{15}$C occupies the weakly bound level $1/2^{+}_{2}$
with the occupation probability of 0.5.
That is, the ground state spin-parity $J^\pi = 1/2^+$, which
is consistent with experimental results as shown in Table~\ref{tab:C15-17-19}.
The main spherical components of the orbital $1/2^{+}_{2}$ are
$1d_{5/2}$ and $2s_{1/2}$ with probability amplitudes
of 0.31 and 0.18.
We define the probability amplitude of a spherical component
in a deformed orbital as the product of the occupation probability of
the orbital and the amplitude of the spherical component
in the deformed wave function.
The $s$-wave component of the weakly bound level $1/2^{+}_{2}$
leads to the neutron halo in $^{15}$C.

\begin{figure}[htb]
\begin{center}
\includegraphics[width=0.56\textwidth]{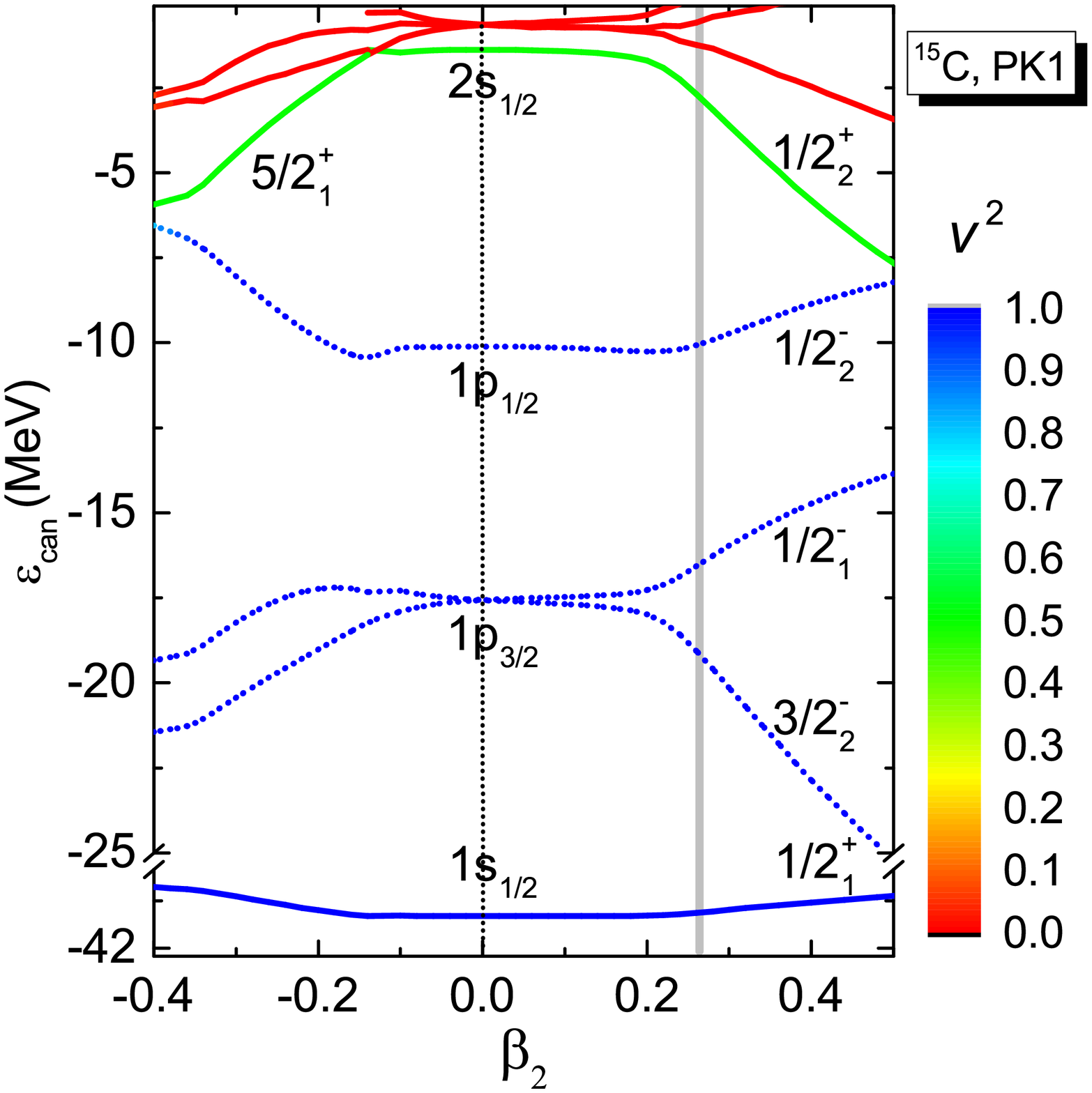}
\end{center}
\caption{(Color online)
Single neutron levels in the canonical basis of $^{15}$C
from deformation constraint DRHBc calculations with PK1.
When $\beta_2 \ne 0$, each level is labeled with $\Omega^\pi_i$
where $\pi$ is the parity,
$\Omega$ is the projection of the angular momentum on the symmetry axis, and
$i$ is used to order the level in each $\Omega^\pi$-block.
The occupation probability $\nu^2$ of each orbital
is represented with different colors.
The ground state is indicated by the gray vertical line.
}
\label{fig:C15_SPL_cst}
\end{figure}

The evolution of single neutron levels with deformation
is studied by making constraint calculations.
The SPLs of $^{16}$C, shown in
Fig.~\ref{figure:SPL-iso}, are used
to guide us to determine the blocked level as a function of $\beta_2$
in constraint calculations of $^{15}$C.
The orbital $5/2_1^+$ crosses $1/2_2^+$
at $\beta_2 \approx -0.06$ in SPLs of $^{16}$C.
In the constraint calculations for $^{15}$C,
the $1/2_2^+$ level is blocked when $\beta_2\ge0$ and
each of the $5/2_1^+$ and $1/2_2^+$ orbitals is blocked respectively when
$\beta_2<0$ in order to get the SPLs
corresponding to the lowest energy state.
Thus obtained SPLs of $^{15}$C are shown in
Fig.~\ref{fig:C15_SPL_cst}.
It is found that there is a crossing between $5/2_1^+$ and $1/2_2^+$
at $\beta_2 \approx -0.15$ in
Fig.~ \ref{fig:C15_SPL_cst}.
This means the valence orbital of neutron is
$1/2^+_2$ when $\beta_2>-0.15$ and
$5/2^+_1$ when $\beta_2<-0.15$.

It is very interesting to see that
in the spherical limit $\beta_2=0$, the neutron orbital
$\nu 2s_{1/2}$ is closer to
$\nu 1p_{1/2}$ than $\nu 1d_{5/2}$ ,
i.e., the inversion of $\nu 2s_{1/2}$ and
$\nu 1d_{5/2}$ appears in DRHBc calculations with PK1.
This inversion plays a
role for the formation of the one-neutron halo
in $^{15}$C, which is similar to the case
in $^{11}$Li \cite{Meng1996_PRL77-3963,Sagawa1992_PLB286-7,
	Borge1997_PRC55-R8,Morrissey1997_NPA627-222}.
In the spherical limit, the valence neutron occupies an $s$-wave orbital,
resulting in a pure $s$-wave halo in $^{15}$C.
However, strong quadrupole correlations drive
$^{15}$C to be prolate with $\beta_2 = 0.26$
in the ground state and mix $sd$ orbitals.
The probability amplitude of $s$-wave component
in the level $1/2^+_2$ is about 0.18 in the ground state,
which is smaller than the one in the spherical case
with the value of 0.5. Furthermore, the $1/2^+_2$ level is more bound in the ground state
compared with $\nu 2s_{1/2}$ in the spherical limit.
This means the deformation effects suppress the halo in $^{15}$C.

\begin{figure}[htb]
\begin{center}
\includegraphics[width=0.5\textwidth]{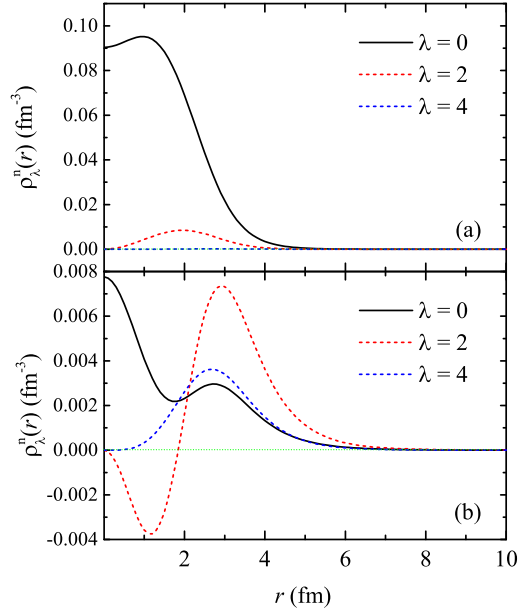}	\end{center}
\caption{(Color online)
The neutron density of $^{15}$C from DRHBc calculations with PK1 decomposed into
spherical ($\lambda=0$), quadrupole ($\lambda=2$), and hexadecapole ($\lambda= 4$)
components for (a) the core and (b) the halo.
}
\label{fig:den-lambda-C15}
\end{figure}

There is a big gap between the level $1/2^+_2$ and $1/2^-_2$
in the ground state of $^{15}$C,
as seen in
both Fig.~\ref{fig:SPL-C15-17-19} and Fig.~\ref{fig:C15_SPL_cst}.
This gap is used to divide the total neutron density into the core
and halo parts. The halo part comes from the weakly bound level
$1/2^+_2$ while the orbital $1/2^-_2$ and those below are deeply
bound and contribute to the core part.
Density profiles of the core
and the halo are presented in
Fig.~\ref{fig:den-C15-PK1}(b) and Fig.~\ref{fig:den-C15-PK1}(c),
respectively. It is obvious that the core of $^{15}$C shows a nearly
spherical shape while the halo is prolate.
This indicates the shape decoupling between the core and halo.
Thus, the shape decoupling between the halo and core is predicted
not only in even-even nuclei with a two-neutron halo
\cite{Zhou2010_PRC82-011301R,Li2012_PRC85-024312,
	Sun2018_PLB785-530},
but also in odd $A$ nuclei with a one-neutron halo.

In Fig.~\ref{fig:den-lambda-C15} the densities of
the core and halo of
$^{15}$C are respectively decomposed into spherical ($\lambda=0$),
quadrupole ($\lambda=2$),
and hexadecapole ($\lambda= 4$) components
[cf.~Eq.~(\ref{eq:expansion})].
In Fig.~\ref{fig:den-lambda-C15}(a),
it can be found that the quadrupole
component of the core is always positive with
very small values compared with the spherical component,
which corresponds to
the nearly spherical (slightly prolate) core of $^{15}$C as
shown in Fig.~\ref{fig:den-C15-PK1}(b).
The nearly spherical core of $^{15}$C is consistent with
the spherical shape of $^{14}$C in Fig.~\ref{fig:density-C}.
The quadrupole component of the halo density is mostly positive
in Fig.~\ref{fig:den-lambda-C15}(b).
This agrees with the prolate shape of the density distribution of
the halo shown in Fig.~\ref{fig:den-C15-PK1}(c).
The shape of the halo originates from the intrinsic
structure of the $1/2^+_2$ level.
The angular distribution of the $s$-wave component is spherical;
the angular distribution of the $1d_{5/2}$ component
in the level $1/2^+_2$ is the mixing of
$|Y_{20}(\Omega)|^2$ and $|Y_{21}(\Omega)|^2$
and it turns out that $|Y_{20}(\Omega)|^2$
has a bigger amplitude.
This means the halo density distribution is prolate because
$|Y_{20}(\Omega)|^2\propto\cos^4\theta$ \cite{Misu1997_NPA614-44}.

\subsection{$^{17}\mathrm{C}$}\label{sub:C17}

\begin{figure}[htb]
\begin{center}
\includegraphics[width=0.6\textwidth]{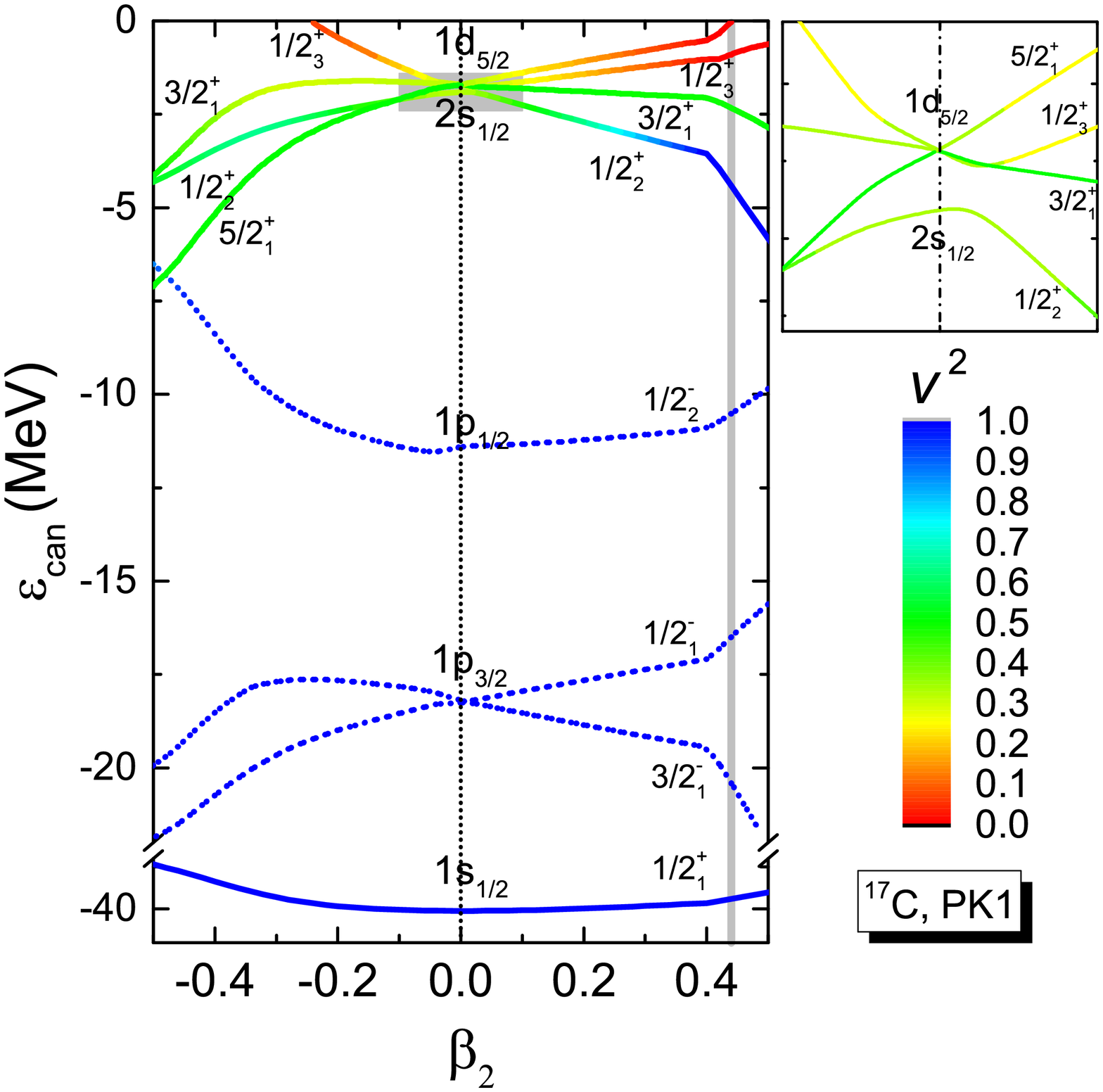}
\end{center}
\caption{(Color online)
Single neutron levels in the canonical basis of $^{17}$C
from deformation constraint DRHBc calculations with PK1.
When $\beta_2 \ne 0$, each level is labeled with $\Omega^\pi_i$
where $\pi$ is the parity,
$\Omega$ is the projection of the angular momentum on the symmetry axis, and
$i$ is used to order the level in each $\Omega^\pi$-block.
The occupation probability $\nu^2$ of each orbital
is represented with different colors.
The ground state is indicated by the gray vertical line.
The shaded region with
$-0.1 \le \beta_2 \le 0.1$ and $-2.3$ MeV
$ \le \varepsilon_{\mathrm{can}} \le -1.3 $ MeV
is enlarged and shown on the right side.
}
\label{fig:C17_SPL_w_block}
\end{figure}

The single neutron levels of $^{17}$C in the ground state
are shown in Fig.~\ref{fig:SPL-C15-17-19}.
The deformation parameter of the ground state is $\beta_2=0.46$.
The valence neutron occupies the $3/2^+_1$ orbital
with the occupation probability of 0.5.
The main spherical components of $3/2^+_1$
are $1d_{3/2}$ and $1d_{5/2}$.
The ground state
spin-parity from DRHBc calculations
agrees with experiments \cite{Maddalena2001_PRC63-024613,
Elekes2005_PLB614-174,Ueno2013_PRC87-034316}.
There is no $s$-wave component in
the valence orbital and
$^{17}$C is not a halo nucleus in its ground sate,
which is contrary to the result from
the SRHFB model~\cite{Lu2013_PRC87-034311}.
Note that the halo structure may appear in low
lying excited states of $^{17}$C, which
has been investigated by using the halo
effective field theory~\cite{Braun2018_EPJA54-196}.

Deformation constraint calculations are performed for this nucleus and
possible blocked levels can be singled out from the
SPLs of $^{18}$C shown in Fig.~\ref{figure:SPL-iso}.
The constraint calculation of $^{17}$C is more
complex compared with $^{15}$C,
because the level density
around $\epsilon_\mathrm{can} \approx -2$ MeV is higher
and there are several crossings among single neutron levels.
The SPLs of $^{17}$C are shown in
Fig.~\ref{fig:C17_SPL_w_block}.
Note that for an odd-$A$ nucleus, the bulk properties 
depend on which orbital is blocked \cite{Zhou2001_PRC63-047305};
so do SPLs in self-consistent calculations.
Therefore one can get several sets of SPLs when various orbitals around
the neutron Fermi surface is blocked.
When getting SPLs shown in Fig.~\ref{fig:C17_SPL_w_block},
the orbital $3/2^+_1$ is blocked for the region $\beta_2>0$
and $5/2_1^+$ is blocked when $\beta_2<0$.
Since both orbitals are from $\nu1d_{5/2}$, the SPLs are continuous
at $\beta_2 = 0$.
There is a sudden change in the slope of each single neutron level
at $\beta_2 \approx 0.40$ where the pairing energy diminishes to zero.
From the enlarged figure shown on the right side of Fig.~\ref{fig:C17_SPL_w_block},
one can notice that the neutron orbital $\nu 2s_{1/2}$ is lower than $\nu 1d_{5/2}$)
in the spherical limit in $^{17}$C with PK1.



\subsection{$^{19}\mathrm{C}$}\label{sub:C19}

\begin{figure}[htb]
\begin{center}
\includegraphics[width=0.56\textwidth]{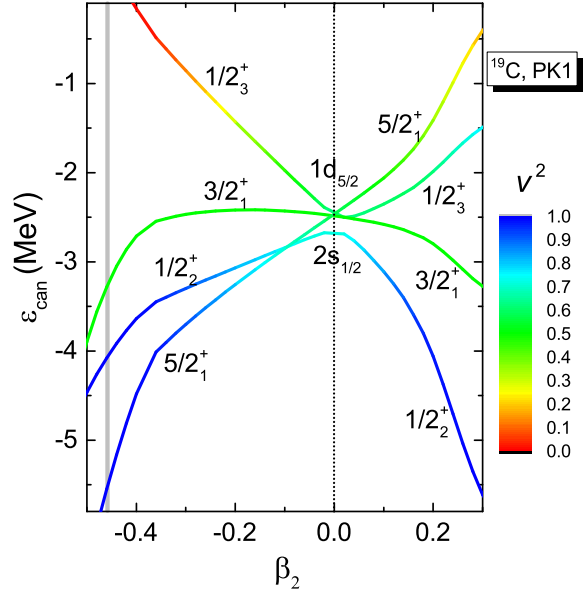}
\end{center}
\caption{(Color online)
Single neutron levels in the $sd$ shell in the canonical basis of $^{19}$C
from deformation constraint DRHBc calculations with PK1.
When $\beta_2 \ne 0$, each level is labeled with $\Omega^\pi_i$
where $\pi$ is the parity,
$\Omega$ is the projection of the angular momentum on the symmetry axis, and
$i$ is used to order the level in each $\Omega^\pi$-block.
The occupation probability $\nu^2$ of each orbital
is represented with different colors.
The ground state is indicated by the gray vertical line.
}
\label{figure:C19_SPL_w_block}
\end{figure}

The intrinsic shape of the ground state of $^{19}$C is oblate
with $\beta_2 = -0.45$ and the rms matter radius is 3.02 fm
in DRHBc calculations with PK1.
The valence orbital is the weakly bound level $3/2^+_1$
with the occupation probability of 0.5.
The main spherical components of the
$3/2^+_1$ level are $d$-waves.
So the ground state of $^{19}$C does not show a halo structure.
This is contrary to the experimental results
\cite{Nakamura1999_PRL83-1112,Maddalena2001_PRC63-024613,
	Elekes2005_PLB614-174,Hwang2017_PLB769-503}.
The SPLs in the ground state of $^{19}$C are shown in
Fig.~\ref{fig:SPL-C15-17-19}.
Quadrupole deformation constraint calculations for
$^{19}$C are carried out and the evolution of SPLs with $\beta_2$
is given in Fig.~\ref{figure:C19_SPL_w_block}.
From the single neutron spectrum of $^{19}$C shown in
Fig.~\ref{figure:C19_SPL_w_block}, one can conclude that
the valence neutron occupies the
$3/2^+_1$ orbital when $\beta_2<0$
and the $1/2^+_3$ orbital when $\beta_2>0$.
It is seen that the inversion of ($\nu 2s_{1/2},
\nu 1d_{5/2}$) in the spherical limit also
exists in $^{19}$C with PK1.


Many theoretical studies support an $s$-wave halo in
$^{19}$C \cite{Ridikas1997_EPL37-385,Kanungo2000_NPA677-171,
Goenuel2001_FBS30-211,Wang2001_NPA691-618,
Acharya2013_NPA913-103,Lu2013_PRC87-034311}.
The ground state spin-parity of $^{19}$C is confirmed to be $J^\pi = 1/2^+$
\cite{Nakamura1999_PRL83-1112,Maddalena2001_PRC63-024613,
Elekes2005_PLB614-174,Satou2008_PLB660-320,Hwang2017_PLB769-503},
implying a prolate intrinsic deformation \cite{Suzuki2003_CM2002-236,Satou2008_PLB660-320}.
Recently $^{19}$C is investigated by using the complex
momentum representation method~\cite{Cao2018_JPG45-085105} and
the authors reproduced the halo structure in
$^{19}$C with the last valence neutron occupying
the $1/2^+$ orbital and the quadrupole deformation $0.2<\beta_2<0.4$.
In DRHBc calculations with PK1, there is a second energy minimum
($E^* = 1.57$ MeV)
in the potential energy surface when the neutron level $1/2^+_3$ is blocked.
The $\beta_2$ value corresponding to this energy minimum is 0.37 which is very
close to the one obtained from deformed Skyrme HF calculations with SGII
\cite{Sagawa2004_PRC70-054316}.
Single neutron levels corresponding to this local energy minimum are given in
Fig.~\ref{fig:SPL-C15-17-19}, labeled as $^{19}\mathrm{C}^*$.
The one-neutron separation energy of $^{19}$C$^*$ is 0.03 MeV.
With a so small one-neutron separation energy,
$^{19}\mathrm{C}^*$ can be regarded as a $^{18}\mathrm{C} +$n system.
The valance orbital is the $1/2^+_3$ level, of which the main
spherical components are $2s_{1/2}$ and $1d_{5/2}$
with probability amplitudes of 0.33 and 0.15.
The halo is thus formed in $^{19}\mathrm{C}^*$, as the result of
the appearance of the $s$-wave component
in the valence orbital $1/2^+_3$.

Due to the inversion of ($\nu 2s_{1/2},\nu 1d_{5/2}$),
the valance neutron occupies $d$-wave orbital and no halo appears
in the spherical limit.
When $\beta_2 > 0$, the level $\nu 1d_{5/2}$ splits into three.
For the highest, $5/2_1^+$, and the lowest, $3/2_1^+$,
the main components are both $\nu 1d_{5/2}$.
Strong quadrupole correlations mix the $sd$ orbitals
and lead to the appearance
of an $s$-wave component in the valence orbital $1/2^+_3$.
Thus, deformation effects are favorable to the formation of the
one-neutron halo in $^{19}\mathrm{C}^*$ under the
inversion of ($\nu 2s_{1/2},\nu 1d_{5/2}$).

\begin{figure}[htb]
\begin{center}
\includegraphics[width=0.7\textwidth]{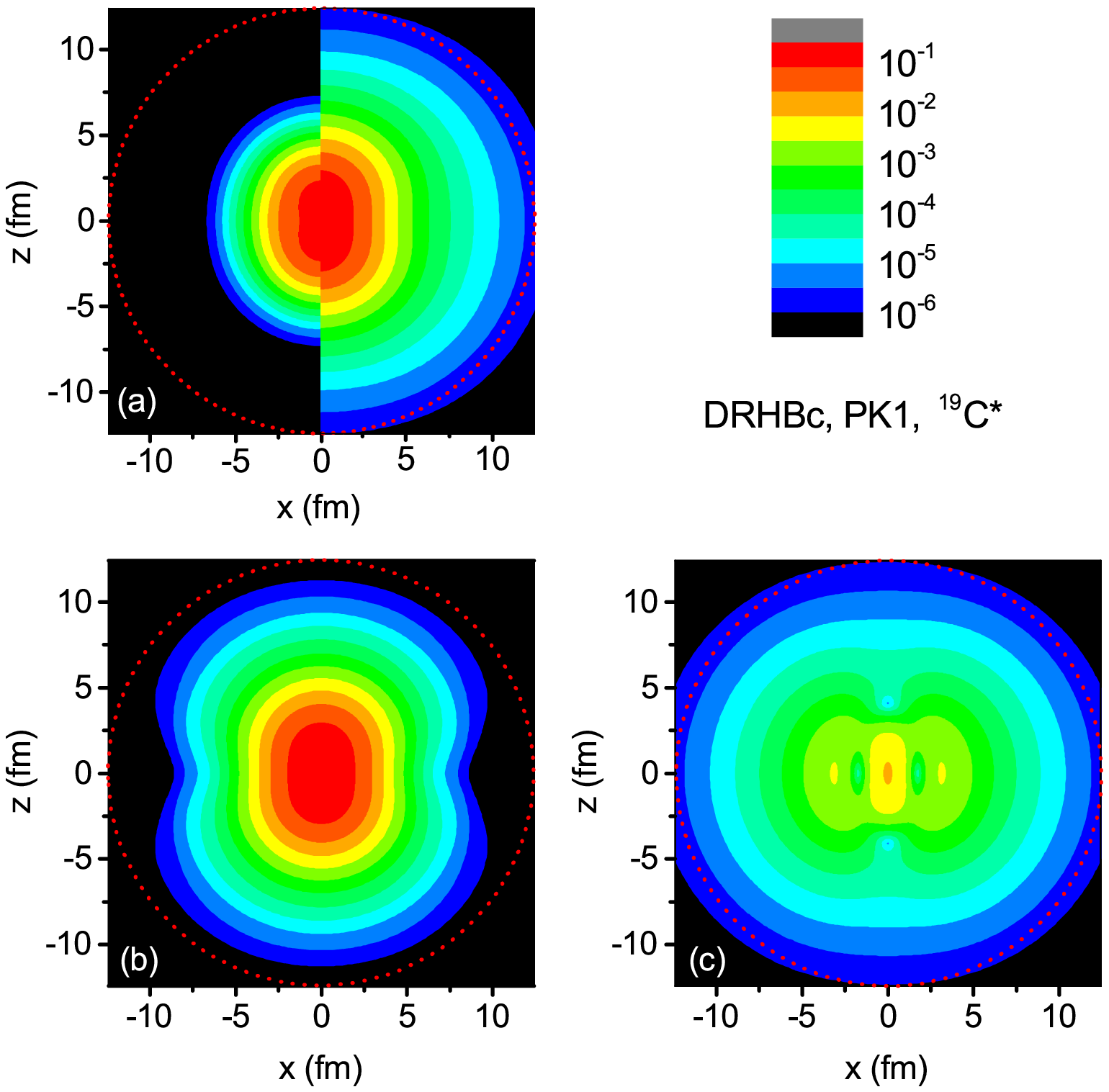}
\end{center}
\caption{(Color online)
Density profiles of $^{19}\mathrm{C}^*$ from DRHBc calculations with PK1.
The $z$-axis is the symmetry axis.
(a) The proton ($x<0$) and neutron ($x>0$) density profiles,
(b) the density profile of the neutron core, and
(c) the density profile of the neutron halo.
In each plot, a dotted circle is drawn to guide the eye.
}
\label{fig:den-C19-PK1}
\end{figure}

\begin{figure}[htb]
\begin{center}
\includegraphics[width=0.5\textwidth]{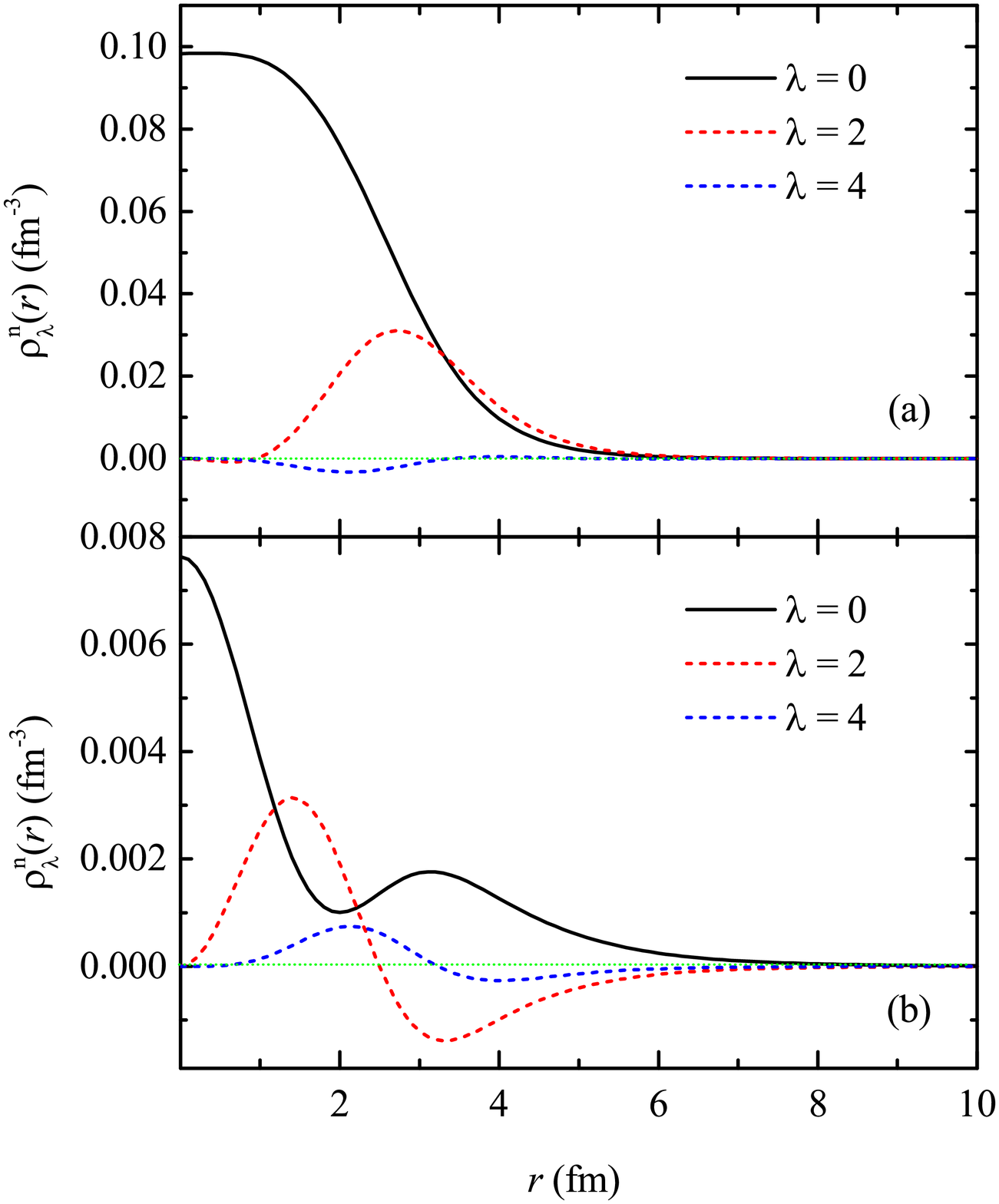}
\end{center}
\caption{(Color online)
The neutron density of $^{19}\mathrm{C}^*$ from DRHBc calculations with PK1 decomposed into
spherical ($\lambda=0$), quadrupole ($\lambda=2$), and hexadecapole ($\lambda= 4$)
components for (a) the core and (b) the halo.
}
\label{fig:den-lambda-C19-PK1}
\end{figure}

Proton and neutron density profiles
for $^{19}\mathrm{C}^*$ are given in
Fig.~\ref{fig:den-C19-PK1}(a).
The total neutron density is divided into the core
and halo parts. The halo part comes from
the weakly bound and valence orbital
$1/2^+_3$. The orbital $3/2^+_1$
and those below are deeply
bound and contribute to the core.
Density profiles of the core
and halo are presented in
Fig.~\ref{fig:den-C19-PK1}(b) and Fig.~\ref{fig:den-C19-PK1}(c),
respectively.
The neutron halo is hinted by the
large extension of the neutron density.
It can be seen that the core of $^{19}\mathrm{C }^*$
shows a prolate shape while the halo
density distribution is slightly oblate.

Spherical ($\lambda=0$),
quadrupole ($\lambda=2$),
and hexadecapole ($\lambda= 4$)
components [cf. Eq.~(\ref{eq:expansion})] of
the core and halo density are shown in
Figs.~\ref{fig:den-lambda-C19-PK1}(a) and (b).
The quadrupole component of the core density is always positive
which is consistent with the prolate shape of
the core density distribution presented in Fig.~\ref{fig:den-C19-PK1}(b).
The quadrupole component of the halo density is positive when $r<2.5$ fm
and negative when $r>2.5$ fm.
Although it looks that the positive amplitude dominates,
the quadrupole moment of the halo part is actually negative due to
the factor $r^2$ in $Q_2 \propto \langle r^2 P_2(\cos\theta) \rangle $.
This indicates an oblate shape of the halo and
the shape decoupling between the core and halo appears in
$^{19}\mathrm{C }^*$.

Up to now, we have discussed the structure of $
^{15,17}$C and $^{19}$C in detail.
We emphasize
that the inversion of ($\nu 2s_{1/2}, \nu 1d_{5/2}$)
together with quadrupole correlations determine
the configuration of the valence neutron orbital
in these three odd $A$ nuclei.
The occupation and closeness to the threshold of
the second and third $1/2^+$ levels, originating from
$\nu2s_{1/2}$ or $\nu1d_{5/2}$,
determine whether or not a halo may appear.
We should also stress that the halo in the
ground state of $^{19}$C is not reproduced
in our DRHBc calculations with PK1,
though a deformed halo appears in the second minimum
of the potential energy curve with $1/2^+_3$ blocked.
This problem should be addressed in future investigations.

\section{Halo in $^{22}$C} \label{sec:C22}

It is believed that the halo in $^{22}$C
is caused by $s$-wave
valence neutrons
\cite{Horiuchi2006_PRC74-034311,
Abu-Ibrahim2008_PRC77-034607,
Abu-Ibrahim2010_PRC81-019901E,
Abu-Ibrahim2009_PRC80-029903E,
Tanaka2010_PRL104-062701,
Coraggio2010_PRC81-064303,
Yamashita2011_PLB697-90,
Fortune2012_PRC85-027303,
Ershov2012_PRC86-034331,Shulgina2018_PRC97-064307,
Kobayashi2012_PRC86-054604,
Frederico2012_PPNP67-939,Hagen2013_EPJA49-118,
Lu2013_PRC87-034311,
Ogata2013_PRC88-024616,Acharya2013_PLB723-196,
Kucuk2014_PRC89-034607,
Hoffman2014_PRC89-061305R,
Inakura2014_PRC89-064316,
Togano2016_PLB761-412,
Ji2016_IJMPE25-1641003,
Pinilla2016_PRC94-024620,
Souza2016_PRC94-064002,
Suzuki2016_PLB753-199,
Souza2016_PLB757-368,
Nagahisa2018_PRC97-054614},
but the probability amplitude of the $s$-wave component
is still not well determined.
In Ref.~\cite{Sun2018_PLB785-530},
the DRHBc theory has been used to study $^{22}$C
and it was revealed that the probability amplitude of
the $s$-wave component is about 0.25.
This value is connected with quadrupole correlations
and the shell evolution
characterized by the inversion of ($\nu 2s_{1/2}$,  $\nu 1d_{5/2}$).
This inversion has been predicted in $A/Z\sim 3$
nuclei \cite{Ozawa2000_PRL84-5493}
and the appearance of shell closures at $N=14$ and $N=16$
is closely related to the competition
between $\nu 2s_{1/2}$ and $\nu 1d_{5/2}$
\cite{Ozawa2000_PRL84-5493,Otsuka2001_PRL87-082502,
	Cortina-Gil2004_PRL93-062501,Brown2005_PRC72-057301,
	Becheva2006_PRL96-012501,Sorlin2008_PPNP61-602,
	Kanungo2009_PRL102-152501,Hoffman2009_PLB672-17,
	Kanungo2013_PST152-014002,Otsuka2020_RMP92-015002}.

\begin{figure}[htb]
\begin{center}
\includegraphics[width=0.7\textwidth]{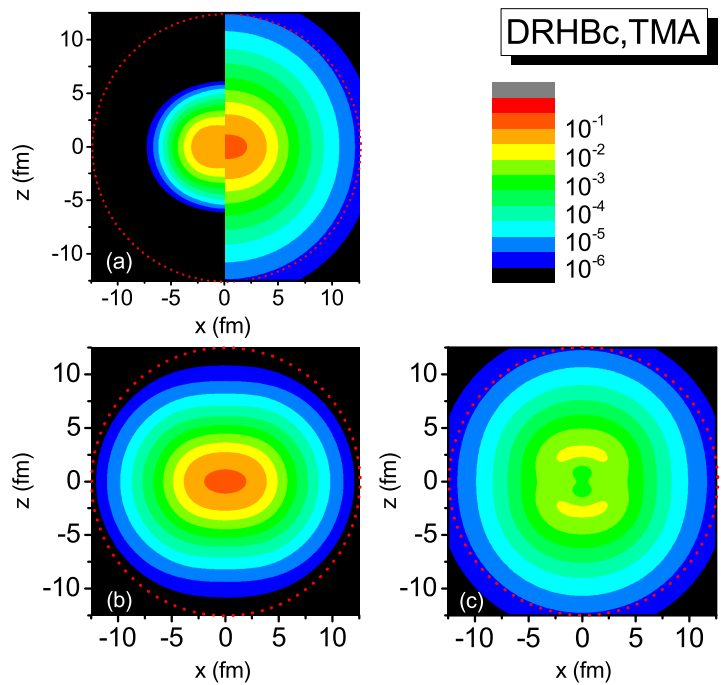}
\end{center}
\caption{(Color online)
Density profiles of $^{22}$C from DRHBc calculations with TMA.
The $z$-axis is the symmetry axis.
(a) The proton ($x<0$) and neutron ($x>0$) density profiles,
(b) the density profile of the neutron core, and
(c) the density profile of the neutron halo.
In each plot, a dotted circle is drawn to guide the eye.
}
\label{fig:den-C22-TMA}
\end{figure}

\begin{figure}[htb]
\begin{center}
\includegraphics[width=0.6\textwidth]{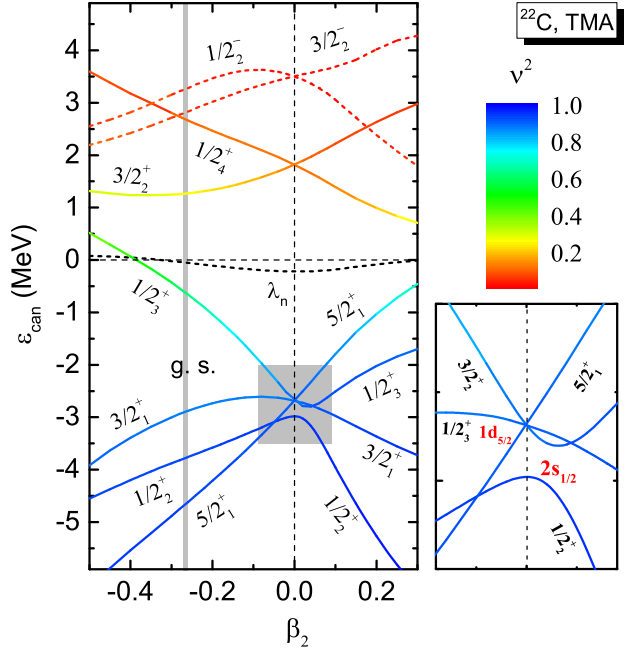}
\end{center}
\caption{(Color online)
Single neutron levels around the Fermi level in the canonical basis of $^{22}$C
from deformation constraint DRHBc calculations with TMA.
Each level is labeled with $\Omega^\pi_i$ where $\pi$ is the parity,
$\Omega$ is the projection of the angular momentum on the symmetry axis, and
$i$ is used to order the level in each $\Omega^\pi$-block.
The Fermi level ($\lambda_n$) is displayed by the black dashed line.
The occupation probability $\nu^2$ of each orbital
is represented with different colors.
The ground state is indicated by the gray vertical line.
The shaded region with
$-0.1 \le \beta_2 \le 0.1$ and $-3.2$ MeV
$ \le \varepsilon_{\mathrm{can}} \le -2.0 $ MeV
is enlarged and shown on the right side.
}
\label{fig:C22-cst-SPL-TMA}
\end{figure}

\begin{figure}[htb]
\begin{center}
\includegraphics[width=0.6\textwidth]{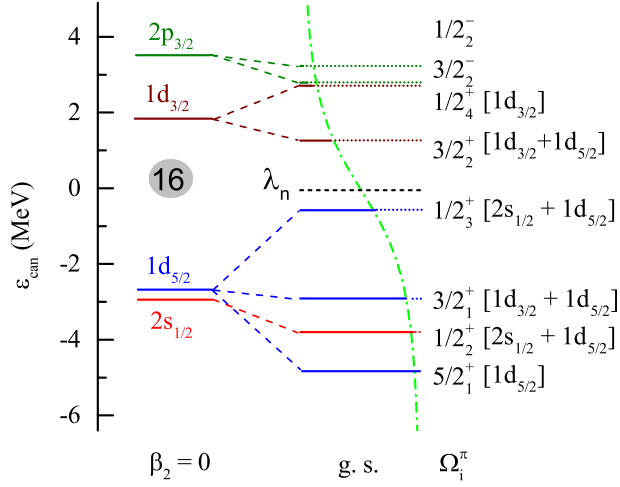}
\end{center}
\caption{(Color online)
Single neutron orbitals around the Fermi level ($\lambda_n$) of $^{22}$C
in the canonical basis in the spherical limit and at the ground state (g.s.)
obtained from DRHBc calculations with TMA.
For the case of the ground state, the length of the solid line is proportional to
the occupation probability of each level calculated from the DRHBc theory.
The dash-dotted line corresponds to the occupation probability calculated
from the BCS formula with an average pairing gap.
Quantum numbers $\Omega^\pi_i$ and the main Woods-Saxon components are
given for orbitals in the $sd$ shell.}
\label{fig:C22-SPL-TMA}
\end{figure}

\begin{figure}[htb]
\begin{center}
\includegraphics[width=0.5\textwidth]{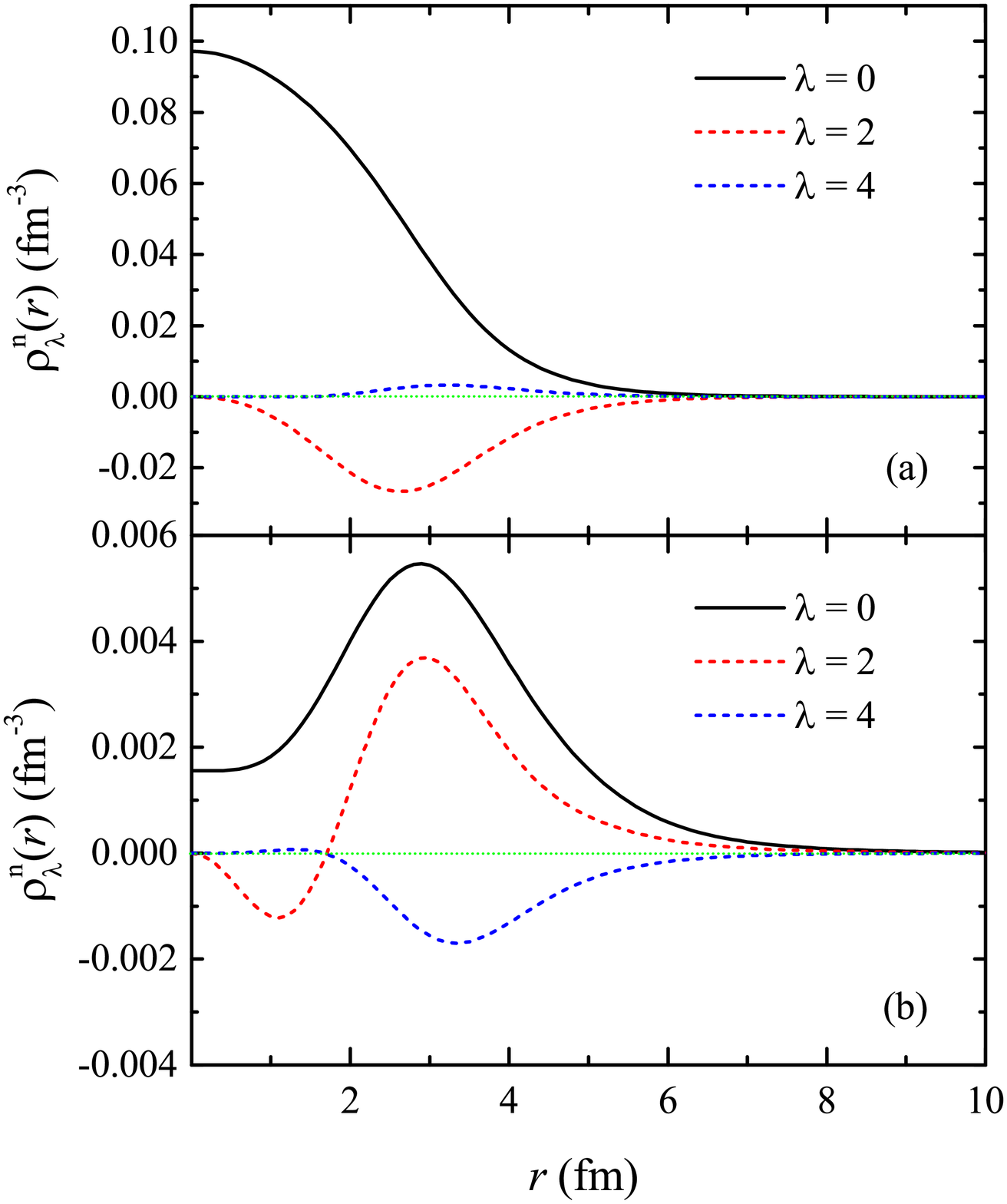}
\end{center}
\caption{(Color online)
The neutron density of $^{22}$C from DRHBc calculations with PK1 decomposed into
spherical ($\lambda=0$), quadrupole ($\lambda=2$), and hexadecapole ($\lambda= 4$)
components for (a) the core and (b) the halo.
}
\label{fig:den-lambda-TMA}
\end{figure}

In Ref.~\cite{Sun2018_PLB785-530}, DRHBc calculations were made
with the effective interaction PK1. To investigate the parameter dependence
and learn more about the results from the DRHBc theory,
we have carried out calculations with
effective interactions NL3 and TMA.
As mentioned earlier,
with NL3 the Borromean feature of $^{22}$C cannot be reproduced.
Next we present the DRHBc results for $^{22}$C
with the parameter set TMA \cite{Toki1995_NPA588-c357}.
The strength of the pairing force $V_0$ is taken as $400$ MeV fm$^{3}$.
Other parameters are the same as those mentioned in Sec.~\ref{sec:bulk}.
The calculated one-neutron separation energy of $^{21}$C is $-0.56$ MeV and
two-neutron separation energy of $^{22}$C is $0.65$ MeV;
i.e., $^{22}$C is a Borromean nucleus.
Considering the large uncertainties in $S_{2n}$ of $^{22}$C,
it should be expected that when $V_0$ is within a range containing
this specific value, the Borromean feature of $^{22}$C can also be reproduced.
The ground state of $^{22}$C has an oblate intrinsic deformation
with $\beta_2 = -0.27$,
which is similar to the result with PK1.
The calculated rms matter radius $R_m$ of $^{22}$C is 3.39 fm with TMA.
This value is consistent
with the data $3.44 \pm 0.08$ fm \cite{Togano2016_PLB761-412}
and $3.38 \pm 0.10$ fm
\cite{Nagahisa2018_PRC97-054614}
but much smaller than the datum $5.4 \pm 0.9$ fm \cite{Tanaka2010_PRL104-062701}.

Next we discuss density profiles
(Fig.~\ref{fig:den-C22-TMA}), single neutron levels
as a function of $\beta_2$ (Fig.~\ref{fig:C22-cst-SPL-TMA}),
single neutron levels corresponding to the ground
state and the spherical shape
(Fig.~\ref{fig:C22-SPL-TMA}),
and different components of the halo and
core densities (Fig.~\ref{fig:den-lambda-TMA}) for $^{22}$C.
It will be seen that these results from DRHBc calculations with TMA
are similar to those with PK1 \cite{Sun2018_PLB785-530}.

The neutron halo in $^{22}$C with TMA can be
clearly seen in Fig.~\ref{fig:den-C22-TMA}(a).
In the single neutron spectrum of $^{22}$C with TMA
shown in Fig.~\ref{fig:C22-cst-SPL-TMA}
and Fig.~\ref{fig:C22-SPL-TMA},
there is a big gap between the two orbitals $1/2^+_3$ and $3/2^+_1$.
Taking advantage of this gap, the total neutron density is decomposed
into the core and halo parts.
The level $3/2^+_1$ and those below it contribute to the core.
The halo part comes from the level $1/2^+_3$ and those above it.
The neutron density distributions of the core and halo are shown
in Figs.~\ref{fig:den-C22-TMA}(b) and (c).
The density distribution of the halo (core) shows a (an) prolate (oblate) shape.
This means the shape decoupling of the core and halo is predicted
in calculations with both PK1 and TMA.
Different components of the core and halo density distributions are presented
in Figs.~\ref{fig:den-lambda-TMA}(a) and (b).
The quadrupole components of the core and halo density,
represented by red dashed line, are smaller than zero and mostly positive,
respectively, being consistent with shapes of density profiles given
in Figs.~\ref{fig:den-C22-TMA}(b) and (c).

The inversion of $\nu 2s_{1/2}$ and $\nu 1d_{5/2}$ is also found
in DRHBc calculations with TMA and
the shell evolution with deformation is similar to that calculated with PK1.
The shell closure at $N=16$ can be seen in the spherical limit
but disappears at the deformed ground state.
Recently, shell model calculations of $^{22}$C and $^{24}$O with continuum considered
\cite{Hu2019_PRC99-061302R}
also predicted a halo in $^{22}$C.
The calculated energy of the first excited state for $^{22}$C is much smaller than
the one in $^{24}$O, hinting the disappearance of the shell closure at $N={16}$ in $^{22}$C.

\section{Summary} \label{sec:summary}

In this work, ground state properties of carbon isotopes are studied in detail
by using the DRHBc theory and compared with experiments.
The DRHBc calculations with the effective interaction PK1 can give a reasonably
good description for bulk properties of nuclei in this isotopic chain.
These carbon isotopes exhibit a variety of intrinsic shapes,
changing abruptly with the neutron number:
spherical $^{13,14}$C,
prolate $^{15,17}$C, and oblate
$^{12,16,18,19,20,22}$C.
Deformation constraint calculations are performed in order to examine the
evolution of SPLs with $\beta_2$.
When $^{15-20,22}$C are constrained to be spherical,
the two neutron orbitals $\nu1d_{5/2}$ and $\nu2s_{1/2}$ are nearly degenerate
but inverted compared with the order of single neutron orbitals in stable nuclei,
leading to the appearance of a shell closure at $N=16$.
When considering deformation effects,
these nuclei are all deformed and the shell closure at $N=16$ disappears.

We have investigated halo structures in $^{15,19,22}$C.
In these three nuclei, halo configurations are closely associated with
the inversion of ($\nu 2s_{1/2},\nu 1d_{5/2}$), deformation effects,
and the shell evolution.
Halos in these nuclei are all caused by a $s$-wave component
in wave functions of valence neutron(s).
$^{15}$C is a one-neuron halo and the intrinsic shape of
its ground is prolate with $\beta_2 =0.26$.
The inversion of ($\nu 2s_{1/2},\nu 1d_{5/2}$) leads to
the formation of the halo because, when constrained to be spherical,
the weakly bound orbital $\nu2s_{1/2}$ is occupied by the valence neutron.
However, the deformation effects suppress the halo in $^{15}$C
due to the following two interwoven reasons:
(i) the valence orbital is $1/2_2^+$ which is more bound than
$\nu2s_{1/2}$ (in the spherical limit);
(ii) the level $1/2_2^+$ is a mixture mainly of $\nu2s_{1/2}$
and $\nu1d_{5/2}$ and not purely of $s$-wave.
It is well accepted that
$^{19}$C is a halo nucleus in its ground state with $J^\pi = 1/2^+$.
The ground state of $^{19}$C in DRHBc calculations
with PK1 shows an oblate shape with the spin-parity of
$3/2^+$ and does not have a halo structure.
Nevertheless, an isomeric state $^{19}\mathrm{C}^*$ with $\beta_2 = 0.37$
corresponding to the second energy minimum of the potential energy surface
with the third 1/2$^+$ neutron level blocked does show an $s$-wave halo.
It should be further investigated why our DRHBc results for $^{19}$C
is at odds with the experiment.
$^{22}$C is reexamined with TMA and similar results have been obtained
as those presented in Ref.~\cite{Sun2018_PLB785-530} with PK1.

For $^{19}$C and $^{22}$C, in the spherical limit, the valence orbital
is $\nu 1d_{5/2}$ and no halo can be formed.
Quadrupole correlations mix $sd$ waves and lead to the appearance of
non-negligible $s$-wave components in the weakly bound orbital
$1/2^+_3$ which is the only valence orbital in $^{19}$C$^*$ and
the most dominant one in $^{22}$C.
This means deformation effects are conducive
to the formation of halos in $^{19}$C$^*$ and $^{22}$C.

We have shown shape decoupling effects in $^{15}$C, $^{19}$C$^*$,
and $^{22}$C with a nearly spherical core but a prolate halo,
a prolate core but a slightly oblate halo, and an oblate core but
a prolate halo, respectively.
We conclude that the shape decoupling between the halo and core exists
not only in even-even nuclei but also in odd $A$ nuclei.

\section*{Acknowledgements}

We thank the DRHBc Mass Table Collaboration, Yu-Ting Rong, and Kun Wang
for helpful discussions.
This work has been supported by
the National Key R\&D Program of China (Grant No. 2018YFA0404402),
the National Natural Science Foundation of China (Grants
No. 11525524, No. 11621131001, No. 11947302, and No. 11961141004),
the Key Research Program of Frontier Sciences of Chinese Academy of Sciences (No. QYZDB-SSWSYS013),
the Strategic Priority Research Program of Chinese Academy of Sciences (Grant No. XDB34010000),
the Inter-Governmental S\&T Coorperation Project between China and Croatia,
and
the IAEA Coordinated Research Project ``F41033''.
The results described in this paper are obtained on
the High-performance Computing Cluster of ITP-CAS and
the ScGrid of the Supercomputing Center, Computer Network Information Center of Chinese Academy of Sciences.



\end{document}